\documentclass[fleqn,usenatbib]{mnras}

\usepackage{newtxtext,newtxmath}

\usepackage[T1]{fontenc}
\usepackage{ae,aecompl}

\usepackage[figure,figure*]{hypcap}
\usepackage{tabularx}
\usepackage{cleveref}
\usepackage{graphicx}
\usepackage{verbatim}

\usepackage{graphicx}	
\usepackage{amsmath}	
\usepackage{amssymb}	
\usepackage[usenames,dvipsnames,svgnames,table]{xcolor}

\newcommand\TP{\mathit{TP}}
\newcommand\FP{\mathit{FP}}
\newcommand\FN{\mathit{FN}}

\newlength\myheight
\newlength\mydepth
\title[Transient Hunting]{Effective Image Differencing with ConvNets for Real-time Transient Hunting}

\author[Nima Sedaghat]{
Nima Sedaghat$^{1}$\thanks{E-mail: nima@cs.uni-freiburg.de (NS)}
 and Ashish Mahabal$^{2}$\thanks{E-mail: aam@astro.caltech.edu (AAM)}
\\
$^{1}$Department of Computer Science, University of Freiburg, Georges-Koehler-Allee 052, 79110 Freiburg, Germany \\
$^{2}$Center for Data Driven Discovery, Caltech, 1200 E California Blvd., Pasadena, CA 91125
}

\date{Accepted XXX. Received YYY; in original form ZZZ}

\pubyear{2017}

\begin{document}
\label{firstpage}
\pagerange{\pageref{firstpage}--\pageref{lastpage}}
\maketitle

\begin{abstract}
Large sky surveys are increasingly relying on image subtraction pipelines for real-time (and archival) transient detection. In this process one has to contend with varying PSF, small brightness variations in many sources, as well as artifacts resulting from saturated stars, and, in general, matching errors. Very often the differencing is done with a reference image that is deeper than individual images and the attendant difference in noise characteristics can also lead to artifacts. We present here a deep-learning approach to transient detection that encapsulates all the steps of a traditional image subtraction pipeline -- image registration, background subtraction, noise removal, psf matching, and subtraction -- into a single real-time convolutional network. Once trained the method works lighteningly fast, and given that it does multiple steps at one go, the advantages for multi-CCD, fast surveys like ZTF and LSST are obvious. 
\end{abstract}

\begin{keywords}
Transient --- Supernova --- Deep Learning --- Artificial Inteligence -- Convolutional Network -- ConvNet
\end{keywords}



\begin{figure*}
  \begin{center}
    \includegraphics[trim={0 0 0 0},clip,width=\linewidth]{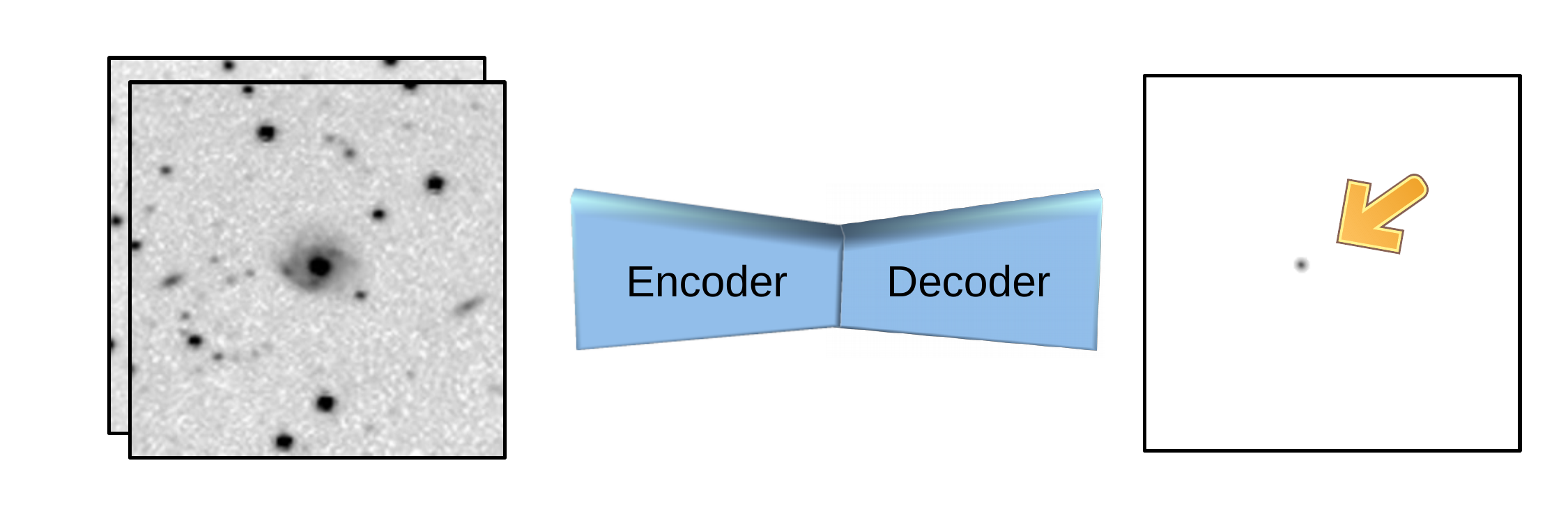}
  \end{center}
  \caption{Our CNN-based encoder-decoder network, TransiNet, produces a difference image without an actual subtraction. It does so through training using a labeled set of transients as the ground-truth.}
  \label{fig:teaser}
\end{figure*}

\section{Introduction}
\label{Sec:intro}

Time-domain studies in optical astronomy have grown rapidly over the last decade with surveys like ASAS-SN 
\citep{ASASSN}, CRTS \citep{Mahabal2011BASI,c6,c7}, Gaia \citep{Gaia}, Palomar-Quest \citep{PQ}, Pan-STARSS
\citep{panstarrs},  PTF \citep{PTF} etc. to name a few. 
With bigger surveys like ZTF \citep{ZTF} and LSST \citep{LSST} around the corner, there is even more interest in the field. 
Besides making available vast sets of objects at different cadences for archival studies, these surveys, combined with fast processing and rapid follow-up capabilities, have opened the doors to an improved understanding of sources that brighten and fade rapidly. The  real-time identification of such sources - called transients - is, in fact, one of the main motivation of such surveys. Examples of transients include extragalactic sources such as the supernovae, and flaring M-dwarf stars within our own Galaxy, to name just two types. The main hurdle is identifying all such varying sources quickly (completeness), and without artifacts (contamination). The identification process is typically done by comparing the latest image (hereafter called the science image), with an older image of the same area of the sky (hereafter called the reference image). The reference image is often deeper so that fainter sources are not mistaken as transients in the science image. Some surveys like CRTS convert the images to a catalog of objects using source extraction software \citep{bertin1996}, and use the catalogs as their discovery domain, comparing brightness of objects detected in the science and reference images. Other surveys like PTF directly difference the reference and science images after proper scaling and look for transients in the difference images.

The reference and science images differ in many ways: (1) changes in the atmosphere mean the way light scatters is different at different times. This is characterized by the point spread function (PSF), (2) the brightness of the sky changes depending on the phase and proximity to the moon, (3) the condition of the sky can be different (e.g. very light cirrus), and (4) the noise and depth (detection limit for faintest sources) are typically different for the two images. As a result, image differencing is non-trivial, and along with real transients come through a large number of artifacts per transient. Eliminating these artifacts has been a bottleneck for past surveys, with humans having been often employed to remove them one by one -- a process called {\it scanning} -- in order to shortlist a set of genuine objects for follow-up using the scarce resources available. Here we present an algorithm based on deep learning that almost completely eliminates artifacts, and is nearly complete (or can be made so) in terms of real objects that it finds. In Sec.~\ref{Sec:relatedWork} we describe prior art for image differencing, and on deep learning in astronomy. In Sec.~\ref{Sec:problemFormulation} we describe the image differencing problem in greater detail, in Sec.~\ref{Sec:method} we present our method and a generative encoder-decoder network -- called {\it TransiNet} hereafter -- based on convolutional networks (ConvNets or CNNS), in Sec.~\ref{Sec:experimets} we detail the experiments we have carried out, and in Sec.~\ref{Sec:discussions} we discuss future directions.

\section{Related Work}
\label{Sec:relatedWork}
For image differencing some of the programs that have been used include 
\citet{Alard-Lupton}, \citet{Bramich2008}, and PTFIDE \citep{PTFIDE}. A recent addition to the list is ZOGY \citep{zogy} which apparently has lower contamination by more than an order of magnitude. It is to be used with the ZTF pipeline and at least in parts of the LSST pipeline. The main task of such an algorithm is to identify new point sources (convolved by the PSF). The problem continues to be challenging because it has to take in to consideration many complicating factors. Besides maximizing real sources found (true positives), generating as clean an image as possible (fewest false positives) is the quantifiable goal. Please refer to \citet{zogy} for greater detail.

Neural networks in their traditional form have been around since as early as 1980s e.g. \cite{rumelhart1986learning} and \cite{lecun_procedure_1985}. 
Such classical architectures have been used in astronomical applications in the past. One famous example is the star-galaxy classifier embedded into the SExtracor package \citep{bertin1996}.

The advent of convolutional neural networks (ConvNets: \cite{lecun1990handwritten,lecun1998gradient}), followed by the advances in parallel computing hardware \citep{raina2009large}, has started a new era in `deep' convolutional networks, specifically in the areas of image processing and computer vision. The applications span from pixel-level tasks such as de-noising to higher-level tasks such as detecting and recognizing multiple objects in a frame. See, e.g. \cite{krizhevsky2012imagenet} and \cite{simonyan2014very}.

Researchers in the area of astrophysics have also very recently started to utilize deep learning-based methods to tackle astronomical problems. Deep Learning has already been used for galaxy classification \citep{DLGal}, supernova classification \citep{c11}, light curve classification \citep{Mahabal2017,Charnock2017}, identifying bars in galaxies \citep{Abraham2017}, separating Near Earth Asteroids from artifacts in images \citep{Bue-inprep}, transient selection post image differencing \citep{Morii2016}, Gravitational Wave transient classification \citep{Mukund2017}, and even classifying noise characteristics \citep{Zevin2017,LVC2017a,George2017}.

One aspect of ConvNets that has not received enough attention in the astrophysical research community, is the ability to generate images as output \citep{rezende2014stochastic,bengio2013generalized}. We provide here such a generative model to tackle the problem of contamination in difference images (see Fig.~\ref{fig:teaser}) and thereby simplify the transient follow-up process.

\section{Problem Formulation}
\label{Sec:problemFormulation}
We cast the transient detection problem as an image generation task. In this approach, the input is composed of a pair of images (generally with different depth, and seeing aka FWHM of the PSF) and the output is an image containing, ideally, only the transient at its correct location and with a proper estimation of the difference in magnitudes. In this work we define a transient as a point source appearing in the second/science input image, and not present in the first/reference image. Such a generative solution as we propose naturally has at its heart registration, noise-removal, sky subtraction, and PSF-matching. 

\begin{figure}
  \begin{center}
    \includegraphics[trim={0 0cm 0 0},clip,width=\linewidth]{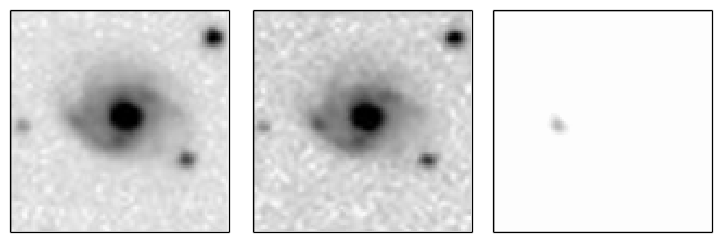}
  \end{center}
  \caption{Examples of the reference (left) and science (center) images. The image on the right is the ground truth output defined for this image pair. It contains the image of a single transient, completely devoid of background and noise. The profile of the transient is the best match to reality our model can produce.}
  \label{fig:formulation}
\end{figure}

In the computer vision literature, this resembles a segmentation task, where one assigns a label to each pixel of an image, e.g. transient vs. non-transient. However our detections include information about the magnitude of the transients and the PSF they are convolved with, in addition to their shape and location. Therefore the pixel-values of the output are real-valued (or are in the same space as the inputs, making it a different problem than a simple segmentation (see Fig.~\ref{fig:formulation}). To this end, we introduce an approach that is based on deep-learning, and train a convolutional neural network (ConvNet) to generate the expected output based on the input image pair.

We formulate the problem as follows. Let us consider $(I_1,I2)$ as the reference-science pair: 
\begin{gather}
\label{eq:I1}
I_1 = I_0\ast\phi_1 + S_1 + N_1\\
\label{eq:I2}
I_2 = (I_0+I_t)\ast\phi_2 + S_2 + N_2
\end{gather}
\noindent where $I_0$ is the underlying unconvolved image of the specific region of the sky; $\phi_1$ and $\phi_2$ represent the PSF models; $S_1$ and $S_2$ are the sky levels and $N_1$ , $N_2$ represent the noise. Note that for the sake of readability, we have illustrated the effect of noise as a simple addition operation. However in reality the noise is `applied' per pixel throughout the workflow.

$I_t$ is the ideal model for the transient, and can be seen as an empty image with an ideal point-source on it.
Based on our formulation of the problem, the answer we seek is $I_t\ast\phi_2$, which represents the image containing the transient, in the same seeing conditions as the science image. This involves PSF matching for taking the first image from $I_0\ast\phi_1$ to $I_0$ and then to $I_0\ast\phi_2$, for the subtraction to work. 

Note that in \cref{eq:I1,eq:I2}, for the sake of clarity, the two images are assumed to be registered. In the real problem that the network is trying to solve, \ref{eq:I2} is replaced by:
\begin{equation} \label{eq:I2_dereg}
I_2 = D\{ (I_0+I_t)\ast\phi_2 \} + S_2 + N_2
\end{equation}
\noindent in which $D\{\}$ represents spatial inconsistency, which in its simplest form consists of one or more of small rotation, translation and scaling.

\begin{figure*}
  \begin{center}
    \includegraphics[width=\linewidth]{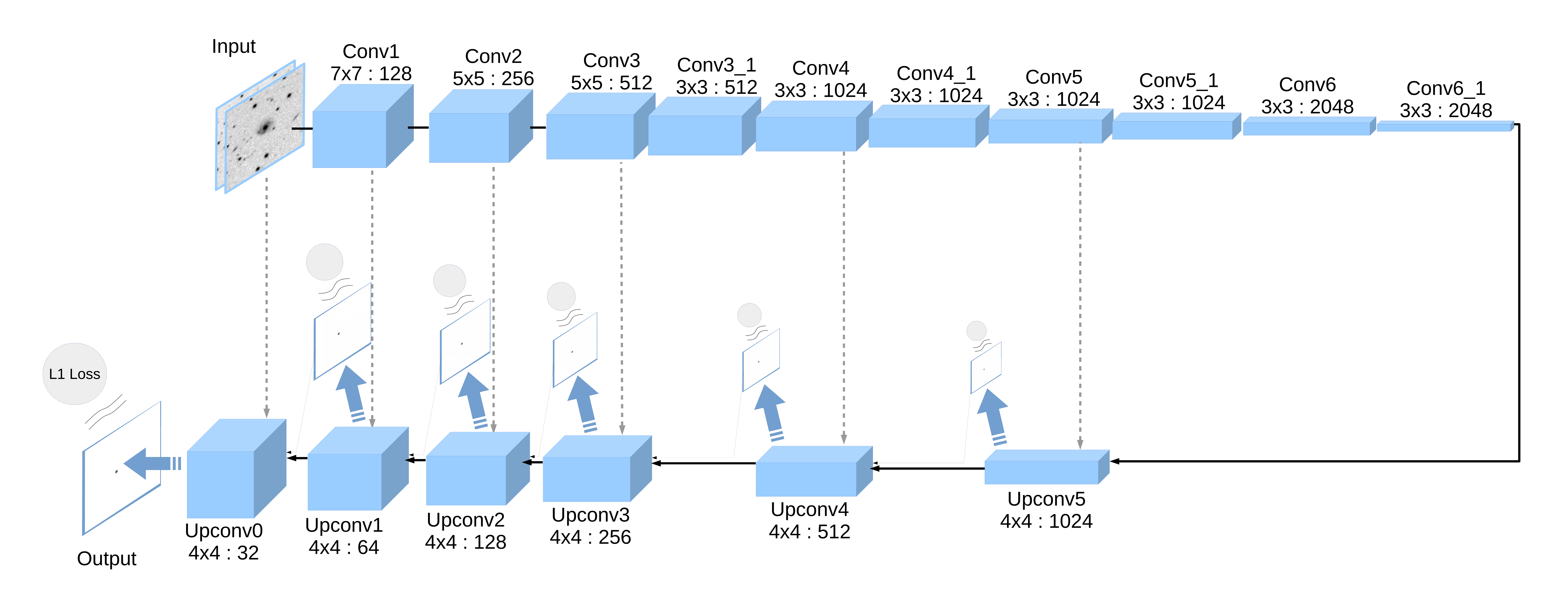}
  \end{center}
  \caption{Our suggested fully convolutional encoder-decoder network architecture. 
  The captions on top/bottom of each layer show the kernel size, followed by the number of feature maps. Each 
  arrow
  represents a convolution layer with a kernel size of 3x3 and stride and padding values of 1, which preserves the spatial dimensions. Dotted lines represent the skip connections. Low-resolution outputs are depicted on top of each up-convolution layer with the corresponding loss. After each (Up-)convolution layer there is a ReLU layer which is not displayed here.}
  \label{fig:architecture}
\end{figure*}

\section{Method}
\label{Sec:method}
We tackle the problem using a deep-learning method, in which an encoder-decoder convolutional neural network is responsible for inferring the desired difference image based on the input pair of images.

\subsection{Network Architecture}

We illustrate TransieNet in Fig.~\ref{fig:architecture}. It is a fully-convolutional encoder-decoder architecture inspired by the one introduced in \cite{NextFlow}. Ten convolutional layers are responsible for the contraction throughout the encoder, and learn features with varying levels of detail in a hierarchical manner. The expansion component of the network consists of 6 up-convolutional layers which decode the learned features, step by step, and generate estimates of the output with different resolutions along the way. 
We compute and back-propagate errors computed based on all different resolutions of the output during training. But in the end and for evaluation purposes, we only consider the full-resolution output. This multi-scale strategy helps the network learn better features with different levels of detail.
We use an L1 loss function at each output:
\begin{equation}
	E = \frac{1}{N} \sum_{n=1}^{N}{|\hat{y_n}-y_n|}
\end{equation}

\noindent where $\hat{y}$ and $y$ represent the prediction and the target (ground truth) respectively, and $N$ is the number of samples in each mini-batch -- see Sec.~\ref{sec:training}. The reason behind the choice of L1 loss over it's more popular counterpart, L2 or Euclidean loss, is that the latter introduces more blur into the output, ruining pixel-level accuracy -- see \cite{NextFlow}, \cite{mathieu_deep_2015}.

\subsection{Data Preparation}
Neural networks are in general data-greedy and require a large training dataset. TransiNet is not an exception and in view of the complexity of the problem -- and equivalently the architecture -- needs a large number of training samples: reference-science image pairs + their corresponding ground-truth images.
Real astronomical image pairs with transients  are not so publicly available. Difficulty of providing proper transient annotations makes them even scarcer. The best one could do is to manually (or semi-automatically) annotate image pairs, and find smart ways to estimate a close-to-correct ground truth image: a clean difference image with background-subtracted gradients. Although as explained in Sec.~\ref{sec:real_data} we implement and prepare such a real training set, it is still too small ($\sim$~200 samples) and if used as is, the network would easily overfit it. 

One solution is to use image augmentation techniques, such as spatial transformations, to virtually increase the size of the training set. This trick, though necessary, is still not sufficient in our case with only a few hundred data samples -- the network eventually discovers common patterns and overfits to the few underlying real scenes.

An alternative solution is to generate a large simulated (aka synthetic) dataset. However, relying only on the synthetic data makes the network learn features based on the characteristics of the simulated examples, making it difficult to transfer the knowledge to the real domain.

Our final solution is to feed the network with both types of data: synthetic samples mixed with real astronomical images of sky with approximate annotations. This along with online augmentation, makes a virtually infinite training set, which has the best of both worlds. We describe details about the datasets used and the training strategy in the following sub-sections.

\subsubsection{Real Data} \label{sec:real_data}

\begin{figure}
  \begin{center}
  	\hspace{0.005\linewidth}
    \includegraphics[width=0.4\linewidth]{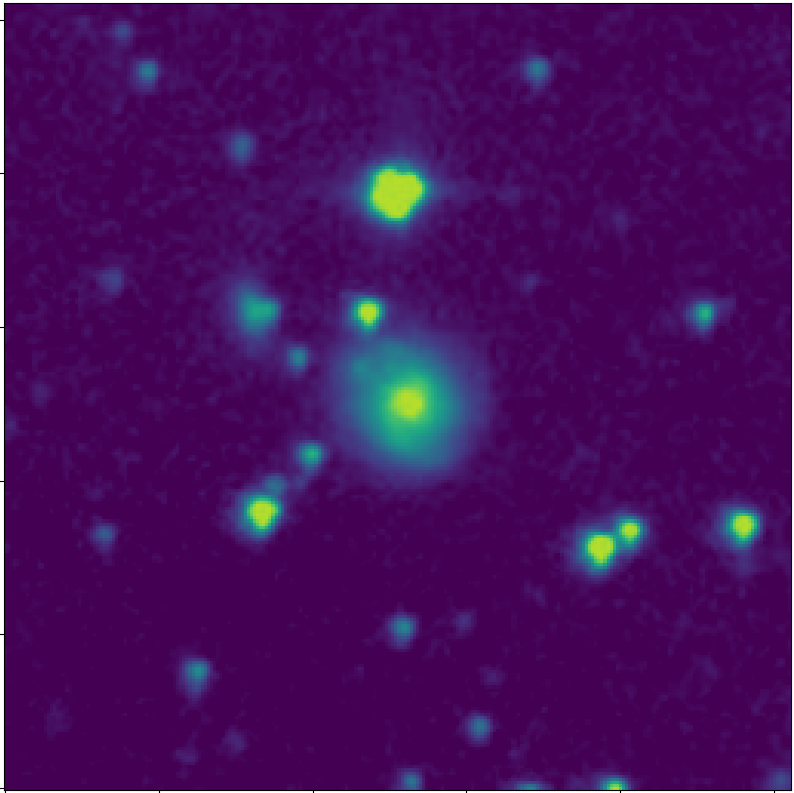}
  	\hspace{0.06\linewidth}
    \includegraphics[width=0.4\linewidth]{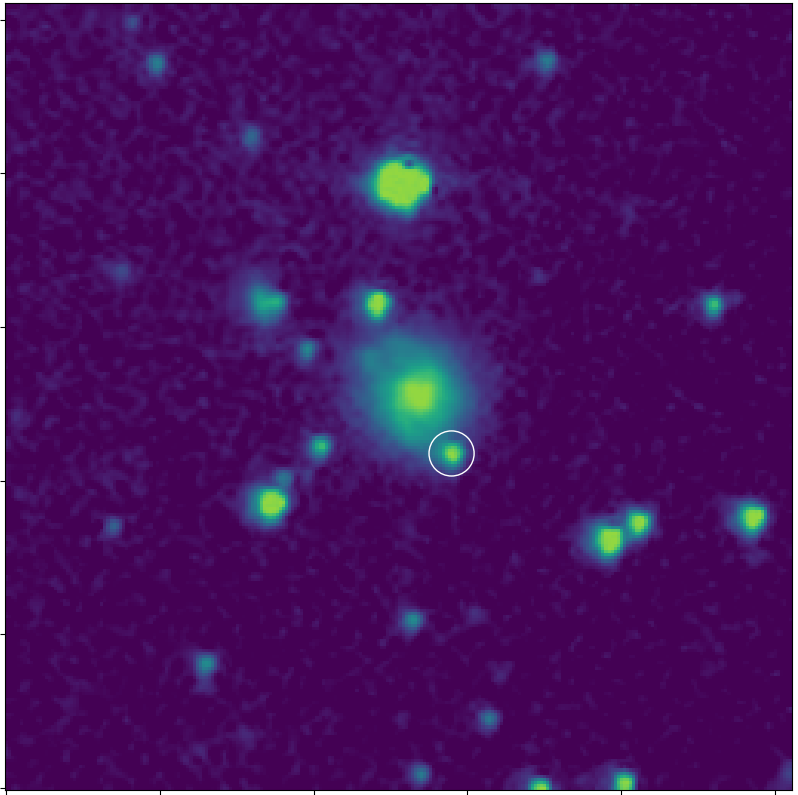}
    \includegraphics[width=\linewidth]{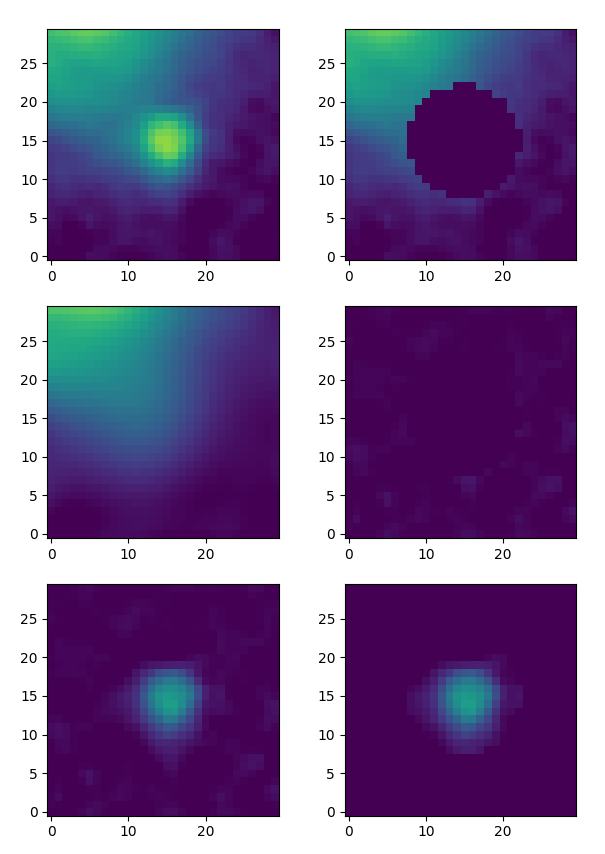}
  \end{center}
  \caption{An exemplar transient annotation case. From top-left to bottom-right the first two images are the input ref/science pair. Number 3 illustrates the $2r\times2r$ neighborhood of the transient and on 4 the user-defined aperture is masked out. Image 5 shows the polynomial model fit to the `masked-neighborhood'. Note that since the blank aperture is excluded from the fitting process, there's no dark region in the results. In number 6 the estimated background is subtracted from the masked neighborhood to form a measure of how well the background has been modeled: the more uniform and dark this image is, the better the polynomial has modeled the background. Finally in number 7 the estimated background is subtracted from the neighborhood and the transient stands out. In number 8 the transient is cropped out of 7 using the user-defined aperture.}
  \label{fig:annot}
\end{figure}
For real examples we make use of data from the Supernova Hunt project \citep{snhunt} of the CRTS survey. In this project image subtraction is performed on pairs of images of galaxies in search of supernovae. While this may bias the project towards finding supernovae rather than generic transients, that should not affect the end result as we mark the transients found, and the ground-truth images contain just the transients. If anything, finding such blended point sources should make finding point sources in the field (i.e. away from other sources) easier. Unlike most other surveys, the CRTS images are obtained without a filter, but that too is not something that directly concerns our method. We gathered 214 pairs of publicly available jpeg images from SN Hunt and split this dataset in to training, validation and test subsets of 102, 26 and 86 members respectively. The reference images are typically made by stacking $\sim~20$ older images of the same area. The science image is a single 30-second exposure. The pixels are $2''.5 \times 2''.5$ and thus comparable or somewhat bigger than the typical PSF. Individual images are $120 x 120$ pixels, and at times not perfectly registered.

To prepare the ground truth, we developed an annotation tool. The user needs to roughly define the location of the transient in the science image, by comparing it with the reference image, and put a circular aperture around it. Then the software models the background and subtracts it from the aperture to provide an estimate of the transient's shape and brightness. Simple annulus-based estimates of the local background \citep{davis1987specifications, howell1989two} or even the recent Aperture Photometry Tool \citep{laher2012aperture},  are not suitable for most of the samples of this dataset since the transients, often supernovae, naturally overlap their host galaxies. Therefore we use a more complex model and fit a polynomial of degree 8 to a square-shaped neighborhood of size $2r \times 2r$ around the aperture, where $r$ is the radius of the user-defined aperture. Note however that the model fitting is performed only after masking out the aperture, to exclude the effect of the transient itself -- the points are literally excluded from model-fitting -- rather than being masked and replaced with a value such as zero. This method works reasonably well even when the local background is complex. Fig.~\ref{fig:annot} illustrates the process.

The annotations on real images are not required to be accurate, as the main responsibility of this dataset is to provide the network with real examples of the sky. This lack of accuracy is compensated by the synthetic samples with precise positions.

\subsubsection{Synthetic Data} \label{sec:synthetic_data}

\begin{figure*}
  \begin{center}
    \includegraphics[width=.95\linewidth]{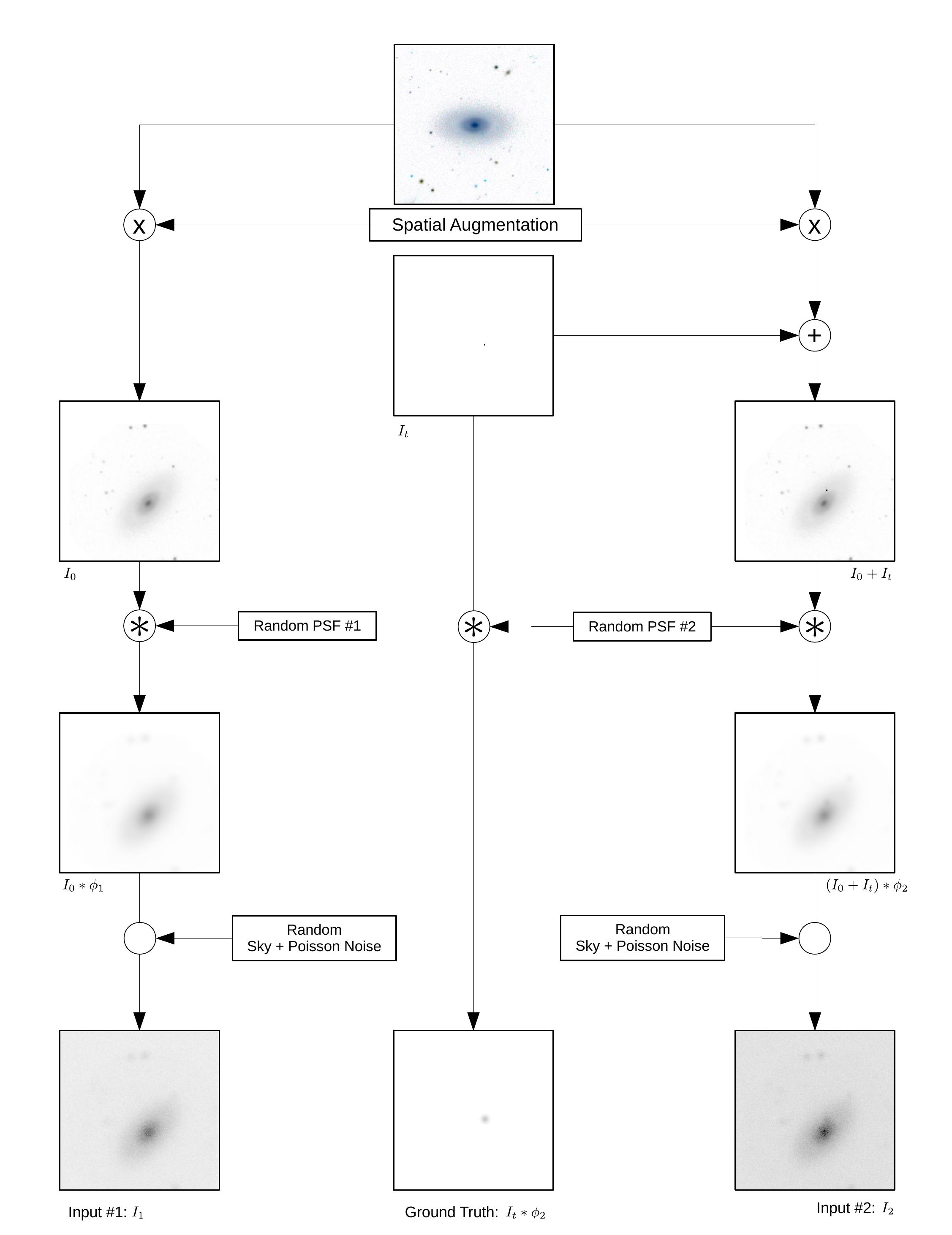}
  \end{center}
  \caption{The synthetic sample generation procedure. The notations used here are described in \Cref{eq:I1,eq:I2}.}
  \label{fig:synthetic}
\end{figure*}

To make close-to-real synthetic training samples, we need realistic background images. Existing simulators, such as Skymaker \citep{bertin2009skymaker}, do not yet provide with a diverse set of galaxy morphologies, and therefore are not suitable for our purpose. Instead, we use images from the Kaggle Galaxy Zoo dataset\footnote{https://www.kaggle.com/c/galaxy-zoo-the-galaxy-challenge}, based on the Galaxy Zoo 2 dataset \citep{Willett2013}, for our simulations. To this end, we pick single images as the background image and create a pair of reference-science images based on it. 

This method also makes us independent of precise physical simulation of the background, allowing us to focus on simulations only at the image level -- even for the `foreground', i.e. transients. This may result in some samples that do not exactly resemble a `normal' astronomical scene, in terms of the magnitude and location of the transient, or the final blur of the objects. But that is even better in a learning-based method, as the network will be trained on a more general set of samples, and less prone to over-fitting to specific types of scenes. 
Fig.~\ref{fig:synthetic} illustrates details of this process.

We first augment the background image using a random spatial transformation:
\begin{gather}
	R \sim U \big( 0,2\pi \big) \\
    T \sim N\Big(\mu=0,\Sigma=
    	\begin{bmatrix}
  		0.03 & 0 \\
  		0    & 0.03
 		\end{bmatrix}
        \Big)
\end{gather}
\noindent here $R$ \& $T$ represent rotation and translation (shift) respectively. $U$ shows a normal distribution and $N$ is a 2D normal distributions. $T$ is then a 2D vector and its values show a translation proportional to the dimensions of the image.

At the next step simulated transients are  added to the science (second) image as ideal point sources, with random locations and magnitudes, to form $I_0 + I_t$. The transient locations are again sampled from a 2D gaussian distribution. The distribution parameters are adjusted such that transients, although scattered all around the image, happen mostly in the vicinity of galaxies at the center of the image, to resemble real supernovae:
\begin{equation}
    \big(X_t,Y_t\big) \sim N\Big(\mu=0,\Sigma=
    	\begin{bmatrix}
  		0.1 & 0 \\
  		0    & 0.1
 		\end{bmatrix}
        \Big)
\end{equation}
In most of our experiments we simulated only a single transient. But in cases where we had more of them, we made sure they were apart from each other by at least half of the bigger dimension of the image. The \textit{amplitude} of the simulated source is also randomly chosen as:
\begin{equation}
    A_t \sim N\big(\mu=10,\sigma=0.3\big)
\end{equation}
\noindent This value, after being convolved with the (sum-normalized) PSF, will constitute the flux of the transient ($F_t$). We can select a specific range of $A_t$ for training -- to fine-tune the network -- based on the range of transients (and their relative brightening) that we expect to find for a given survey.

The two images are then convolved with \textit{different} gaussian PSFs, generated based on random kernel parameters, with a random eccentricity, limited by a user-defined maximum:
\begin{gather}
	\sigma_{\phi,x} \sim U \big(\sigma_{\phi,m},\sigma_{\phi,M}\big) \label{sigma_phi_x}
    \\
	\sigma_{\phi,y} = \sigma_{\phi_1,x} \sqrt{1-{ecc}^2} \label{sigma_phi_x}
	\\
	{ecc} \sim N\big(\mu=0,\sigma={ecc}_{max}\big)   
\end{gather}

\noindent where $\sigma_{\phi,x}$ and $\sigma_{\phi,y}$ are the standard deviations of the 2D gaussian function along x and y directions respectively and $[\sigma_{\phi,m},\sigma_{\phi,M}]$ is the range from which $\sigma_{\phi,x}$ is uniformly sampled.
The PSF is then rotated using a random value, $\theta_\phi$, uniformly sampled from $[0,2\pi]$. This should also help catch asteroids that would leave a very short streak.

Precisely modeling the difference between reference and science images, and adjusting the PSF parameters' distributions accordingly, would be achievable. However, as stated before, we prefer to keep the training samples as general as possible. Therefore in our experiments $[\sigma_{\phi,m},\sigma_{\phi,M}]$ is set to $[2,5]$ for both images. These numbers are larger than typically encountered, and real images should fare better. The ${ecc}_{max}$ value is set to 0.4 and 0.6 for reference and science images, respectively, to model the more isotropically blurred seeing of reference images.

The sky and noise levels are different for the reference and science images. We choose to model these difference in our simulations since in contrast to the previous parameters, ignoring them would make learning easier for the network -- and that's exactly what we want to avoid. We model the sky level, $S$, as a constant value, add it to the image and only after that `apply' the Poisson noise to every pixel:
\begin{equation}
	{I_n(x,y)} = poisson(\lambda = I(x,y)+S)
\end{equation}

\noindent where $poisson$ is a function returning a sample from a Poisson distribution with the given $\lambda$ parameter, $S$ is the sky model, and $I_n$ is the noisy version of input $I$.

Then we perform a pairwise augmentation (rotation, scaling and translation), such that the two images are not perfectly registered. This forces the network to also learn the task of registration on the fly. 

The ground truth image is then formed by convolving the ideal transient's image, $I_t$, with the same PSF as applied to the science image. No constant sky value or noise are applied to this image. This way the network learns to predict transient locations and their magnitudes in the same seeing conditions as the science image, in addition to noise removal and sky subtraction.

\begin{figure*}
	\begin{center}
    	\newcolumntype{Y}{>{\small\centering\arraybackslash}X}
        \begin{tabularx}{\linewidth}{YYYYY}
            {Reference} & {Science} & {ZOGY - D} & {ZOGY - (Scorr>5$\sigma$)} & {TransiNet} \\
        \end{tabularx}
        \end{center}

  \begin{center}
    \includegraphics[width=.95\linewidth]{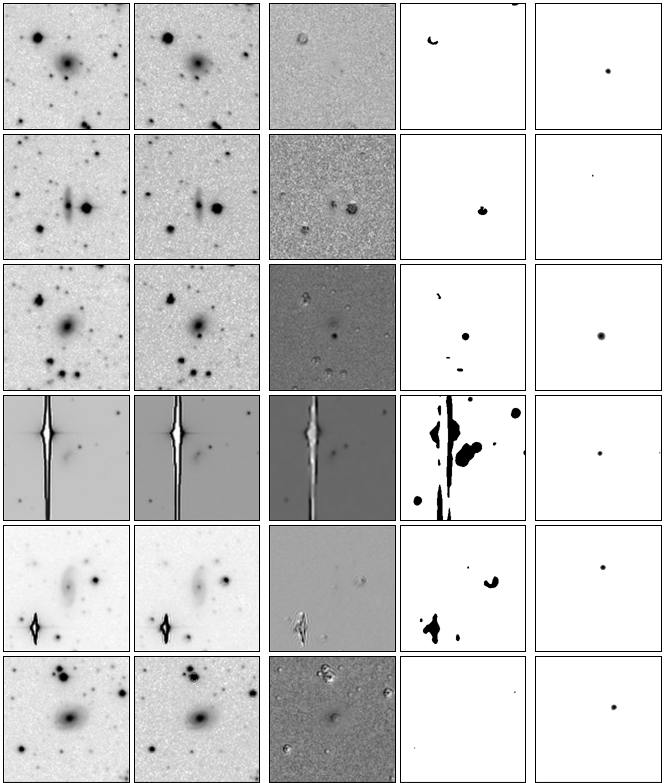}
  \end{center}
  \caption{Image subtraction examples using ZOGY and TransiNet for a set of CRTS Supernova Hunt images. The first column has the deep reference images, second column contains the science images which have a transient source and are a shallower version of the reference images. The third column contains the ZOGY D images, and the fourth  has the ZOGY Scorr images i.e. ``the matched filter difference image corrected for
source noise and astrometric noise" \citep{zogy}. The fifth column has the thresholded versions of ZOGY SCORR, as recommended in that paper. The sixth column shows the difference image obtained using TransiNet.
All images are mapped to the [0,1] range of pixel values, with a gamma correction on the last column for illustration purposes. TransiNet has a better detection accuracy, and is also robust against noise and artifacts. It is possible that ZOGY could be tuned to perform better, and on a different dataset provide superior results - the reason for the comparison here is to simply show that TransiNet does very well.}
  \label{fig:results}
\end{figure*}


\subsection{Training Details} \label{sec:training} We have two networks, one generic, and the other specialized to the SN Hunt dataset. Each shares the 90K images from Kaggle zoo, with transients inserted to create {\it synthetic} science images. The {\it real} data currently is just the $\sim100$ SN Hunt image pairs earmarked for training. The zoo and SN Hunt images are further rotated, shifted, and scaled to augment the dataset, and also make the network more robust. Training is done in small batches of 16. We use ADAM for optimization using the Caffe framework \citep{jia2014caffe}. First 90K iterations are common to both networks. Images used are $140\times140$. Then, for the generic network fine tuning is done using batches that contain 12 CRTS and 4 zoo images. The image size here is $256\times256$ (with images scaled where needed), and the number of iterations is 50K. We put 12 real images and 4 synthetic images in each batch during this second round of training to prevent over-fitting to the small-sized real dataset. We start with a learning rate of $3e^{-4}$ and drop it in the second round by a factor of 0.3 every 20K iterations.
Running on an NVIDIA GTX 1070 along with 16 CPU cores, the whole training process takes a day and half to complete. 

For the specialized SN Hunt network, on the other hand, the entire fine-tuning is done using 8K iterations on CRTS images (we emphasize that we are still using just the $\sim100$ image-pairs, modified in many ways). That way it is better at recognizing transients in real data. 

\subsection{The Attention Trick} 
In this specific type of application, the target images  mainly consist of black regions (i.e. zero intensity pixels), and non-zero regions take up only a small number of pixels. Therefore mere use of a simple L1 loss does not generate and propagate big enough error values back to the network, when the network has learned to remove the noise and generates blank images. So the network spends too long a time focusing on generating blank images instead of the desired output, and in some cases fails to even converge.
The trick we use to get around this issue is to conditionally boost the error on the interesting regions. The realization of this idea is to simply apply the mapping $[0,1] \rightarrow [0,K]$ on the ground-truth pixel values. K represents the boosting factor and we set it to 100 in our experiments. This effectively boosts the error in non-zero regions of the target, virtually increasing the learning rate for those regions only. The output of the network is later downscaled to lie in the normal range.
Note that increasing the total learning rate is not an alternative solution, as the network would go unstable and would not even converge.

\section{Experiments and Results}
\label{Sec:experimets}
We have run TransiNet on samples from CRTS SN Hunt and the Kaggle zoo dataset. The network weights take up about 2GB of memory. Once read, on the NVIDIA GTX 1070 the code runs fast: 39ms per sample, which can be reduced to 14ms if samples are passed to the network as batches of 10. The numbers were calculated by running tests on 10000 images three times. Fig.~\ref{fig:results} depicts samples from running TransiNet on the CRTS test subset. The advantage of TransiNet is that the ``image differencing'' produces a noiseless image ideally consisting of just the transient, and thus robust to artifacts and removes the need for human scanners. 

\begin{figure}
  \begin{center}
    \fbox{\includegraphics[width=0.3\linewidth]{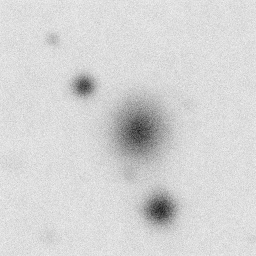}}
    \fbox{\includegraphics[width=0.3\linewidth]{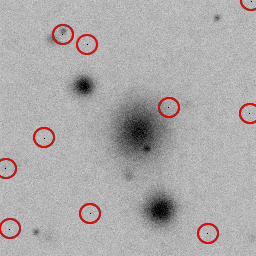}}
    \fbox{\includegraphics[width=0.3\linewidth]{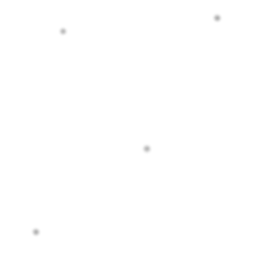}}
  \end{center}
  \caption{An exemplar multi-transient case from the zoo dataset. The reference image (left), science image (middle) with 10 single-pixel Cosmic Ray events, indicated by red circles, and four transients, and the network prediction (right) with all transients detected cleanly, and all CRs rejected.}
  \label{fig:multi}
\end{figure}

\begin{figure}
  \begin{center}
    \fbox{\includegraphics[width=0.3\linewidth]{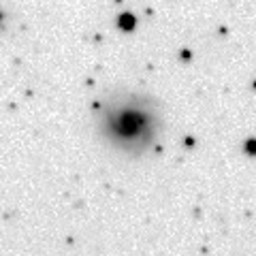}}
    \fbox{\includegraphics[width=0.3\linewidth]{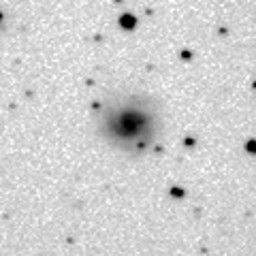}}
    \fbox{\includegraphics[width=0.3\linewidth]{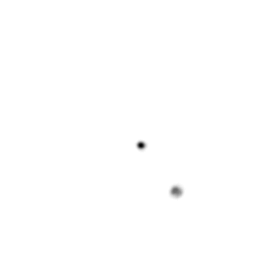}}
  \end{center}
  \caption{An exemplar multi-transient case from the CRTS SN Hunt dataset. The science image (middle) has two transients, and the network prediction (right) finds them both though never trained explicitly to look for multiple transients.}
  \label{fig:multiSNH}
\end{figure}

With increasing CCD sizes it is much more likely than not that there will be multiple transients in a single image.
Since the SN Hunt images, or the zoo images used, rarely have multiple transients, the networks may not be ideal when looking for such cases. However, because of the way the network is trained - with the output as pure PSF-like transients, it is capable of finding multiple transients though we did not train it explicitly with such cases. This is demonstrated in
Fig.~\ref{fig:multi} which depicts an exemplar sample from the zoo subset. Here we introduced four transients, and all were correctly located. Another side effect - a good one - is that the network rejects non-PSF like additions, including Cosmic Rays. In addition to the four transients, we had also inserted 10 single pixel Cosmic Rays in the science image shown in Fig.~\ref{fig:multi} and all were rejected.
An example from the SN Hunt set is shown in Fig.~\ref{fig:multiSNH}
which happens to have two astrophysical objects -- the second is likely an asteroid. Here too, the network has detected both transients. Locating new asteroids is as useful as locating transients to help make the asteroid catalog more complete for future linking and position predicting.

\subsection{Quantitative Evaluation}
We provide below quantitative evaluations of TransiNet performance. 

\subsubsection{Precision-Recall Curve}
Precision-recall curves are the de-facto evaluation tool for detectors. 
They capture $TruePositives$ (TP, or `hits', the number of correctly detected objects), $FalsePositives$ (FP, or `false alarms'), and $FalseNegatives$ (FN, or `misses', the number of missed real objects) for all possible thresholds. This allows users to either set a fixed threshold, or a dynamic threshold (e.g. $5\sigma$ above the background level, or $70\%$ of the max in a difference image etc.) 
\begin{gather} 
	Precision = \frac{TruePositive} {TruePositive + FalsePositive} \label{eq:precision_recall_eq1_1}
    \\
	Recall = \frac{TruePositive} {TruePositive + FalseNegative} \label{eq:precision_recall_eq1_2}
\end{gather}

\paragraph*{Low-SNR Detections \& Blank Outputs}
The output of TransiNet is an image with real-valued pixels. Therefore each pixel is more likely to contain a non-zero real value, even in the `dark' regions of the image, or when there is no transient to detect at all.
Thus, we consider low-SNR detection images as blank images. The outputs of the network (detection images) which have a standard deviation ($\sigma$) lower than $0.001$ were marked as blank images during our experiments and unconsidered thereafter.

\begin{figure*}
	\begin{center}
    	\newcolumntype{Y}{>{\small\centering\arraybackslash}X}
        \begin{tabularx}{\linewidth}{YYYYYYYYY}
            {Ref. Image} & {Sci. Image} & {Raw Det. Image} & {$\tau=\frac{\sigma}{10000}$} & {$\tau=\frac{\sigma}{1000}$} & {$\tau=\frac{\sigma}{100}$} & {$\tau=\frac{\sigma}{10}$} & {$\tau=\sigma$} & {$\tau=10\sigma$}\\ 
        \end{tabularx}
        \end{center}

  \begin{center}
    \fbox{\fbox{\includegraphics[width=0.10\linewidth]{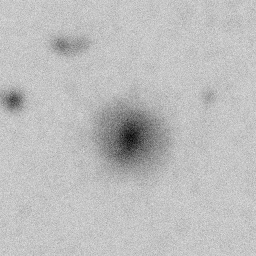}}}
    \fbox{\fbox{\includegraphics[width=0.10\linewidth]{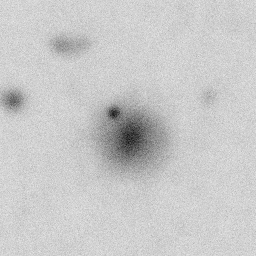}}}
    \fbox{\fbox{\includegraphics[width=0.10\linewidth]{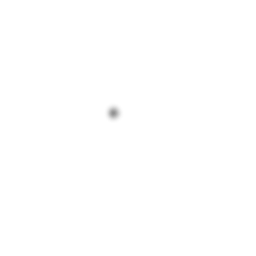}}}
    \hspace{10px}
    \fbox{\includegraphics[width=0.10\linewidth]{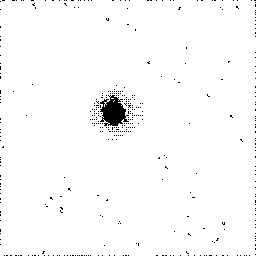}}
    \fbox{\includegraphics[width=0.10\linewidth]{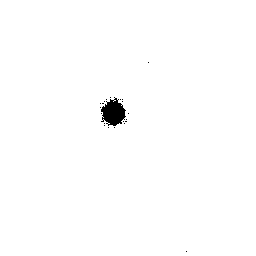}}
    \fbox{\includegraphics[width=0.10\linewidth]{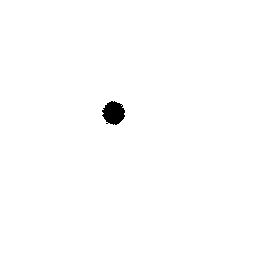}}
    \fbox{\includegraphics[width=0.10\linewidth]{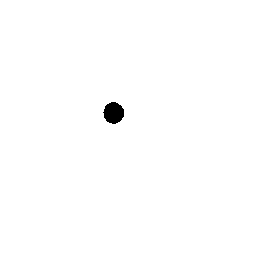}}
    \fbox{\includegraphics[width=0.10\linewidth]{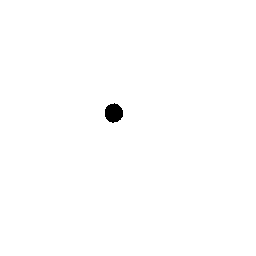}}
    \fbox{\includegraphics[width=0.10\linewidth]{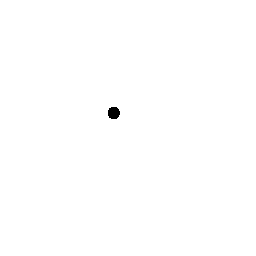}}

    \fbox{\fbox{\includegraphics[width=0.10\linewidth]{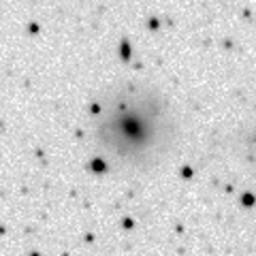}}}
    \fbox{\fbox{\includegraphics[width=0.10\linewidth]{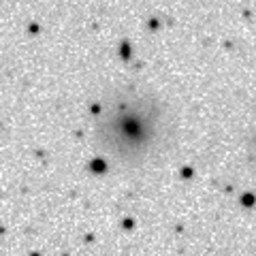}}}
    \fbox{\fbox{\includegraphics[width=0.10\linewidth]{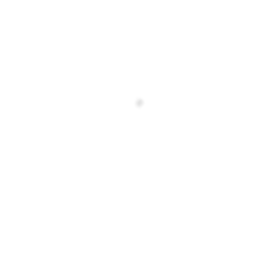}}}
    \hspace{10px}
    \fbox{\includegraphics[width=0.10\linewidth]{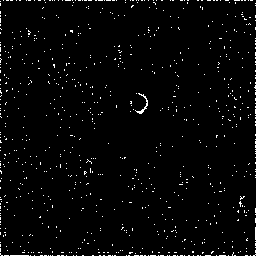}}
    \fbox{\includegraphics[width=0.10\linewidth]{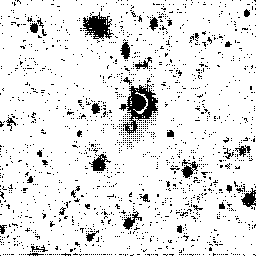}}
    \fbox{\includegraphics[width=0.10\linewidth]{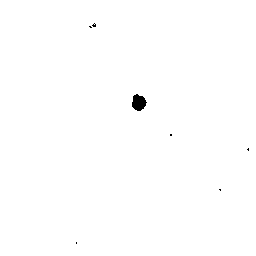}}
    \fbox{\includegraphics[width=0.10\linewidth]{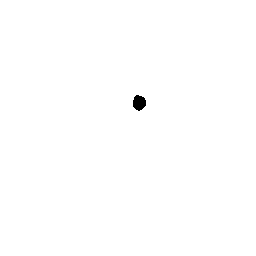}}
    \fbox{\includegraphics[width=0.10\linewidth]{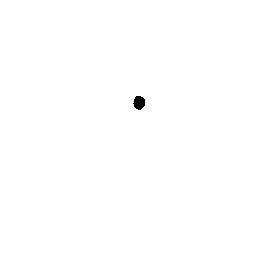}}
    \fbox{\includegraphics[width=0.10\linewidth]{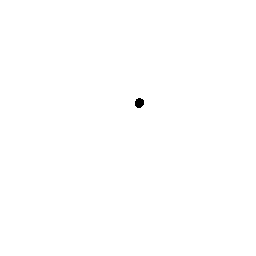}}
    
  \end{center}
  \caption{Visualization of the thresholding process used for generation of precision-recall curves. Each row illustrates exemplar levels of thresholding of a single detection image: first row is chosen from the synthetic subset and the second row from CRTS. Outputs of the network are normally quite clean, and contaminants practically appear only after taking the threshold down below the noise level. This is particularly visible in the second row, where the transient has been of a low magnitude, and so the detection image has a low standard deviation ($\sigma$). Thus $\frac{\sigma}{100}$ is still too low and below noise level.}
  \label{fig:thresholding}
\end{figure*}

\paragraph*{Binariziation and Counting of Objects}
Evaluations at a series of thresholds is the essence of a precision-recall curve and helps reveal low SNR contaminants, while digging for higher completeness (see Fig.~\ref{fig:thresholding}).

The thresholding
\begin{equation}
	\hat{Y}_{ij} = 
    \begin{cases} 
      0 & \hat{y}_{ij} < \tau \\
      1 & \hat{y}_{ij} \geq \tau 
   \end{cases}
\end{equation}

\noindent results in the binary image, $\hat{Y}$, on which we obtain `connected' regions to count detected objects with full connectivity \citep{fiorio1996two,wu2005optimizing}. For this specific kind of evaluation, we also convert the ground truth image ($y$) to a similar binary-valued image, $Y$, using a fixed threshold.

\begin{figure}[]
  \begin{center}
    \includegraphics[width=0.48\linewidth]{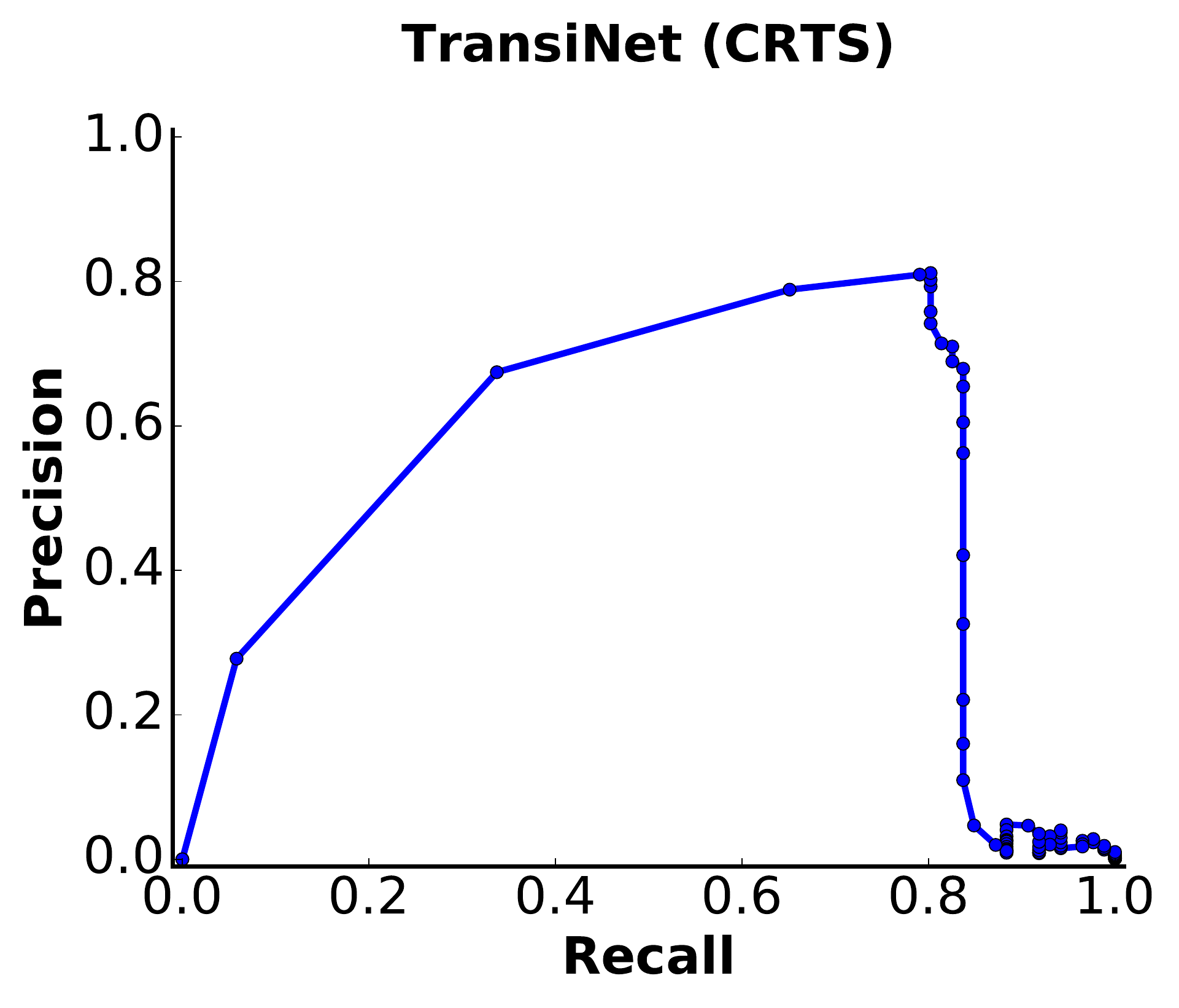}
    \includegraphics[width=0.48\linewidth]{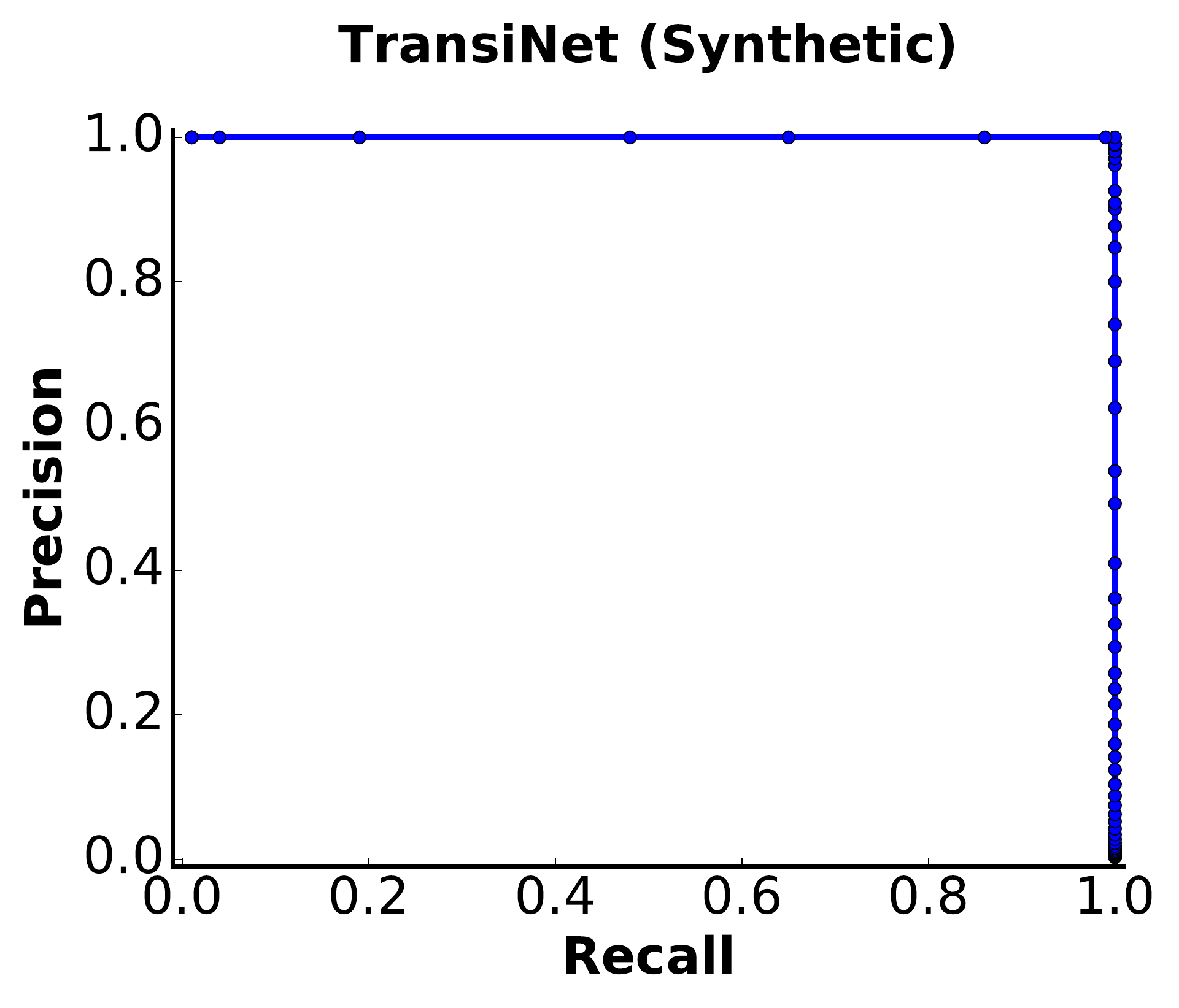}
    \includegraphics[width=0.48\linewidth]{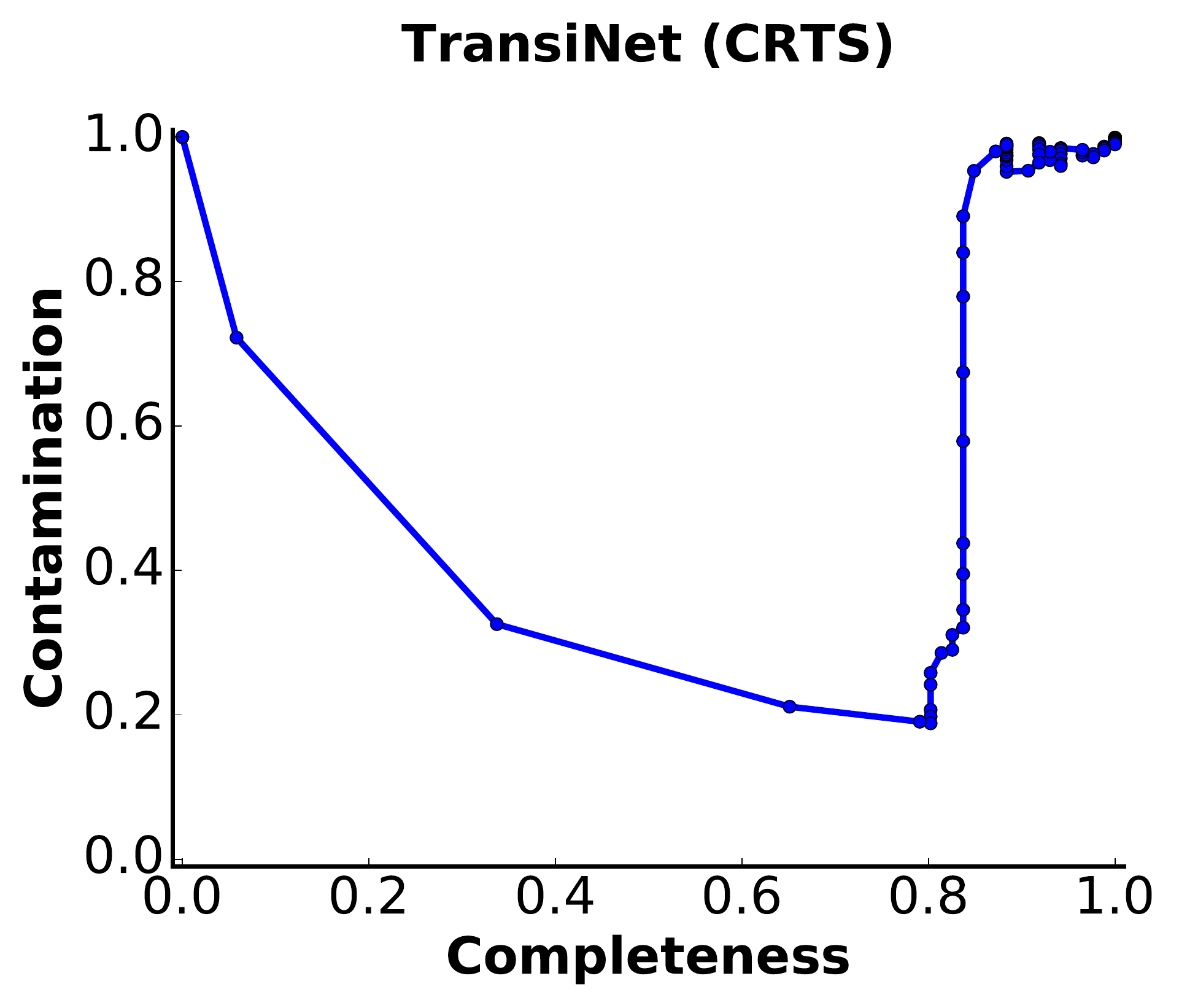}
    \includegraphics[width=0.48\linewidth]{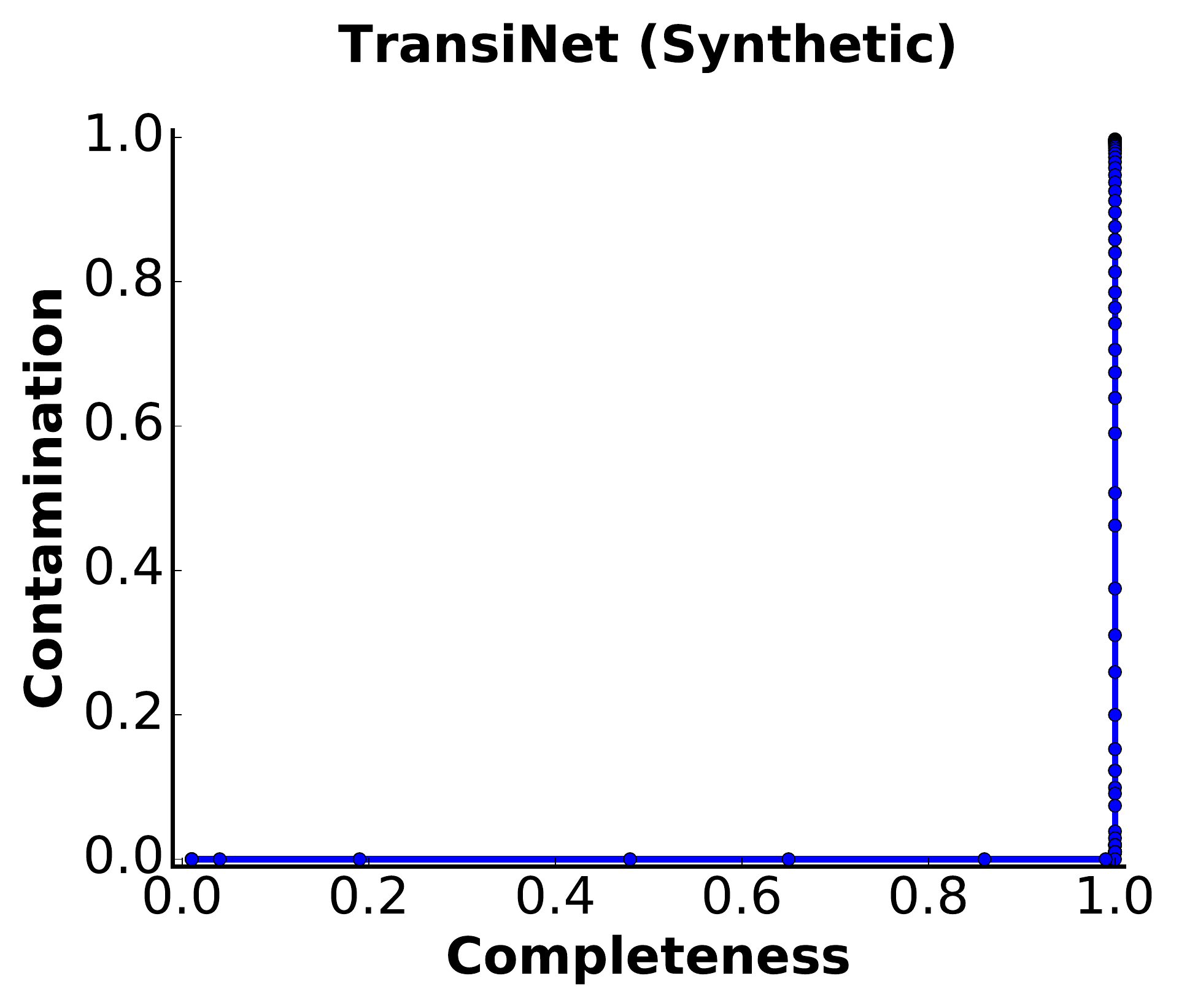}
    \includegraphics[width=0.48\linewidth]{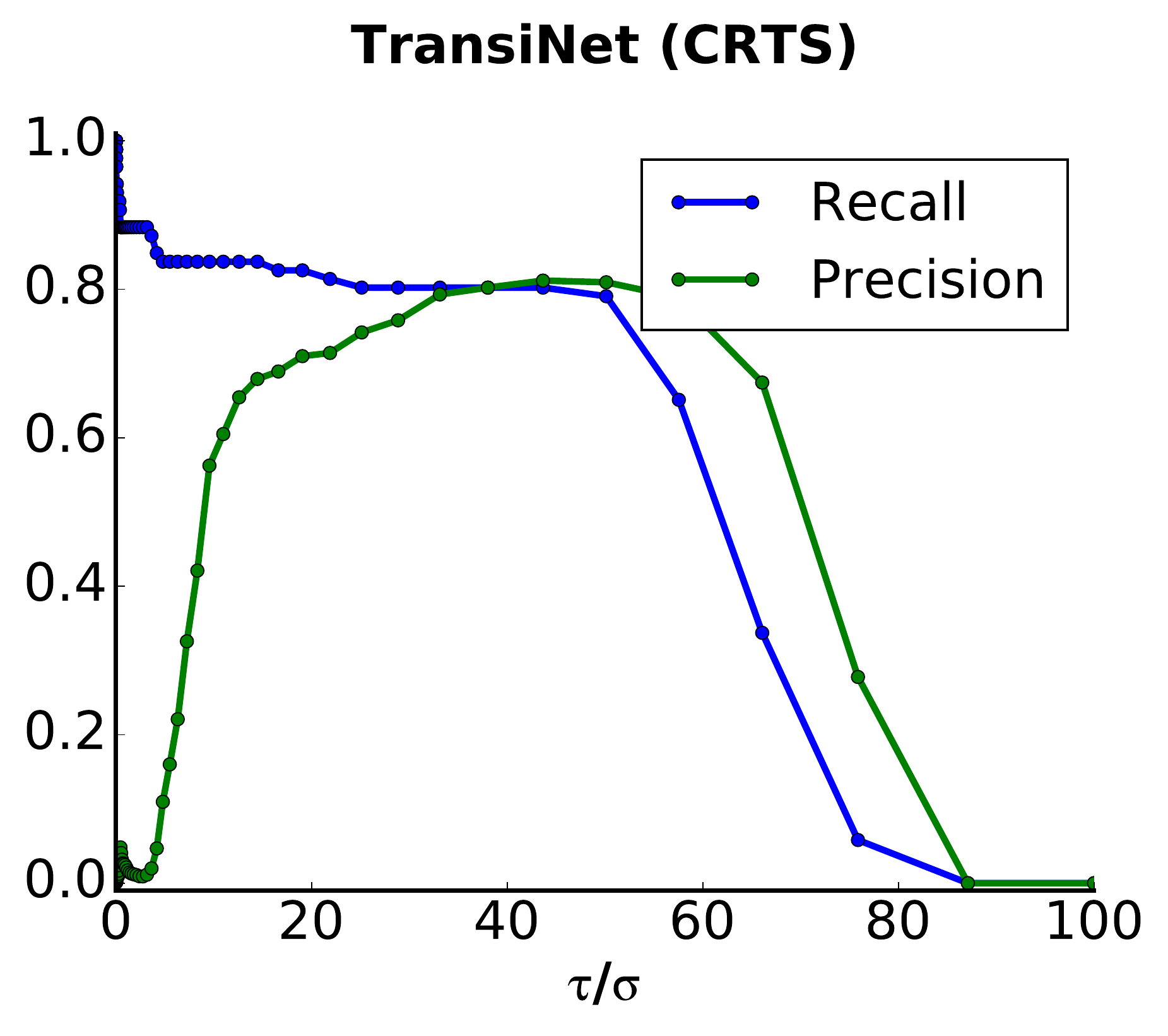}
    \includegraphics[width=0.48\linewidth]{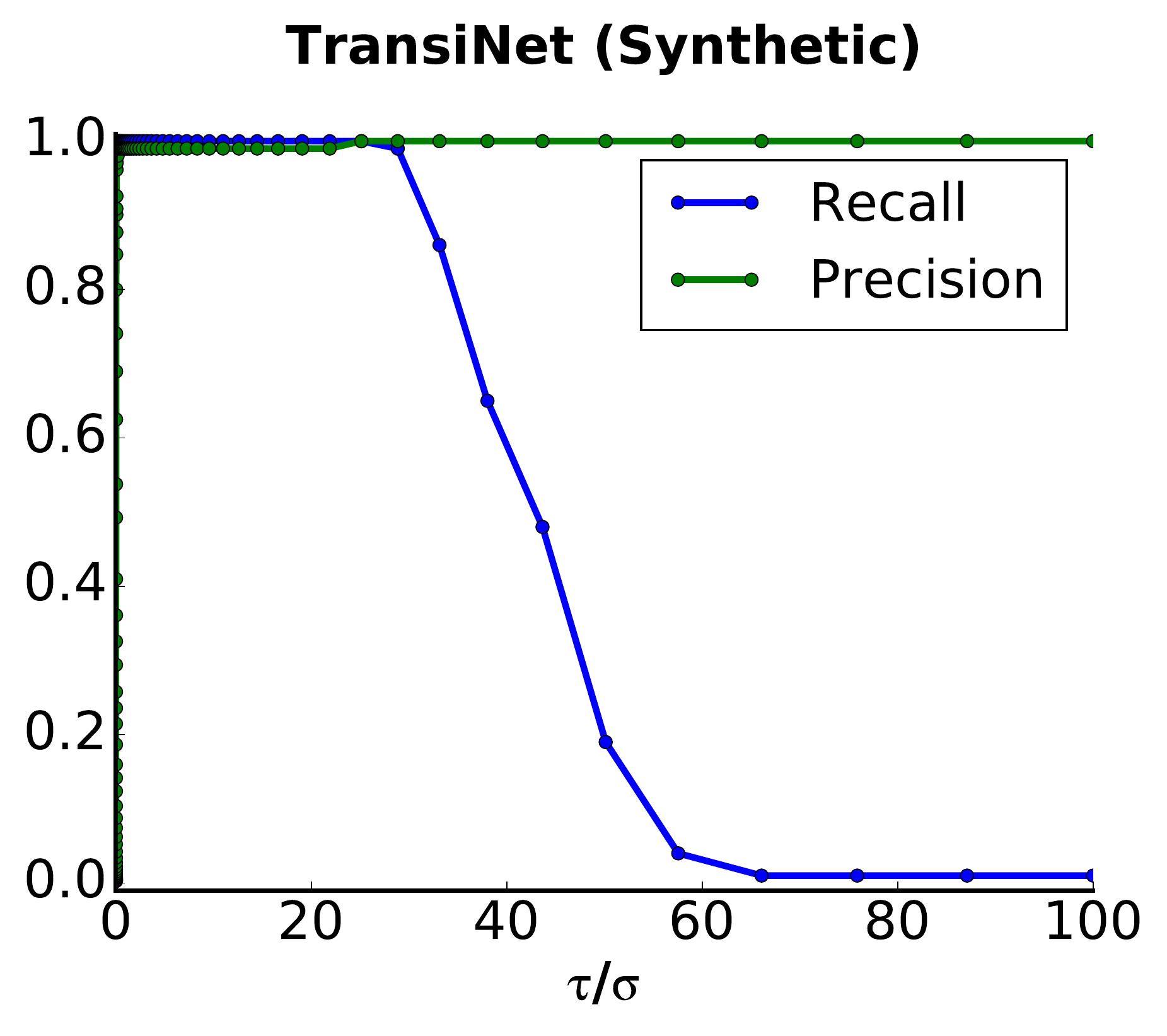}
    \includegraphics[width=0.48\linewidth]{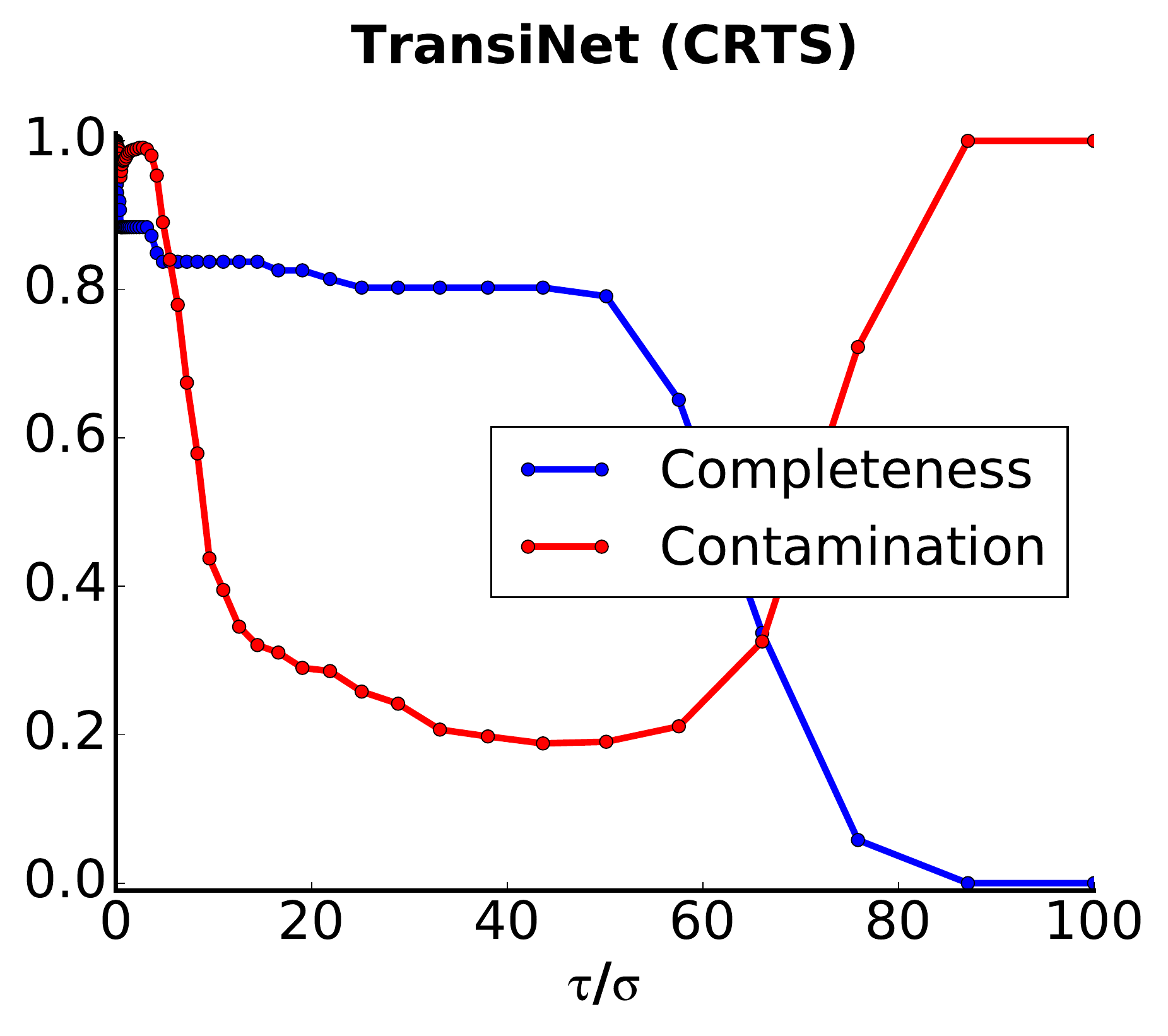}
    \includegraphics[width=0.48\linewidth]{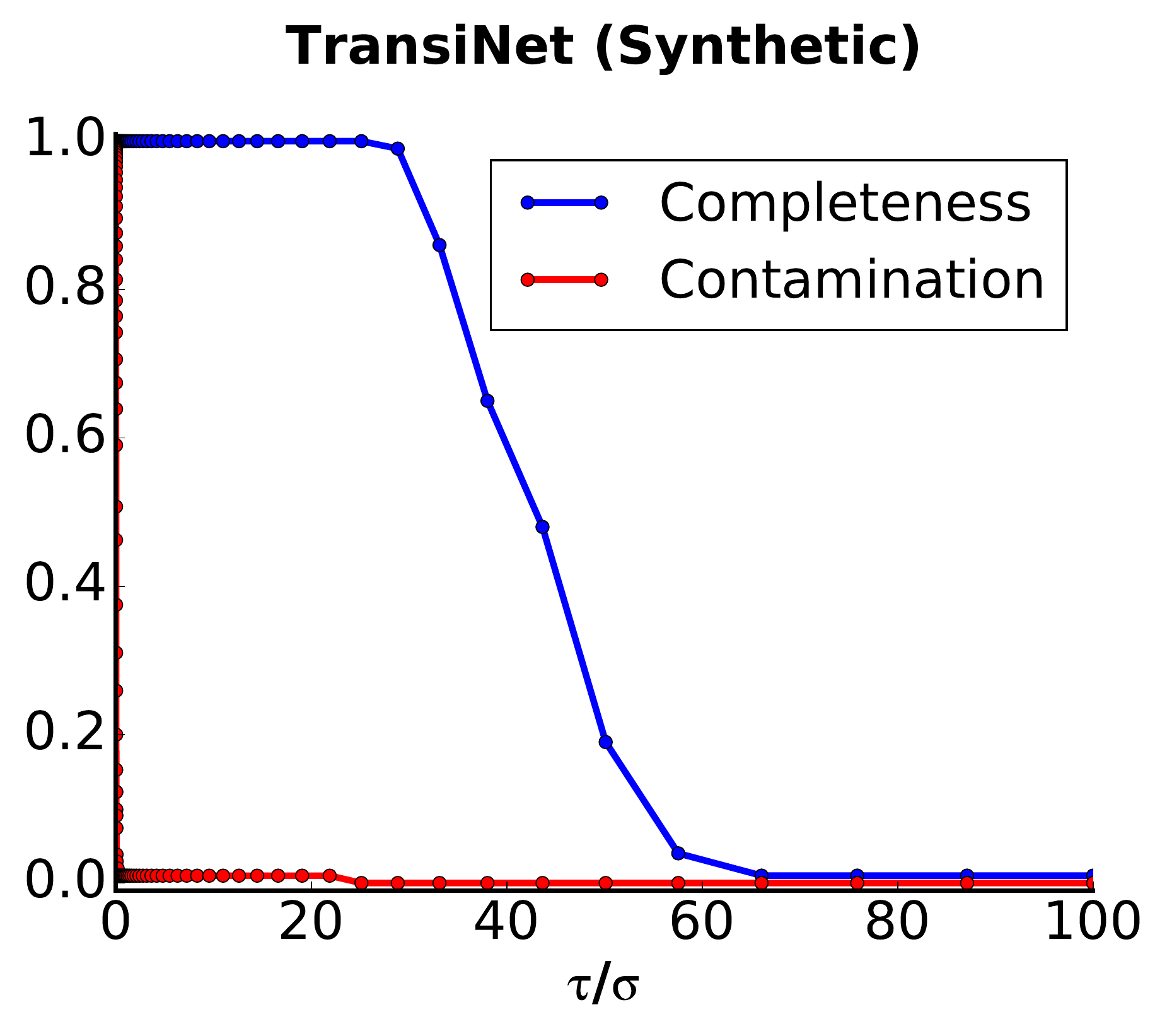}
    
  \end{center}
  \caption{Plots showing precision-recall (row 1), completeness-contamination (row 2), and their dependence on threshold (rows 3 and 4) for TransiNet before the blanking is done to remove ultra low-SNR detections (see text). The two columns show CRTS (left) and synthetic (right) subsets. For CRTS a threshold can be picked where 80\% transients are detected with little contamination. Not unexpectedly, the performance is better for the synthetic images.}
  \label{fig:precision-recall-preblanking}
\end{figure}

\begin{figure}[]
  \begin{center}
    \includegraphics[width=0.48\linewidth]{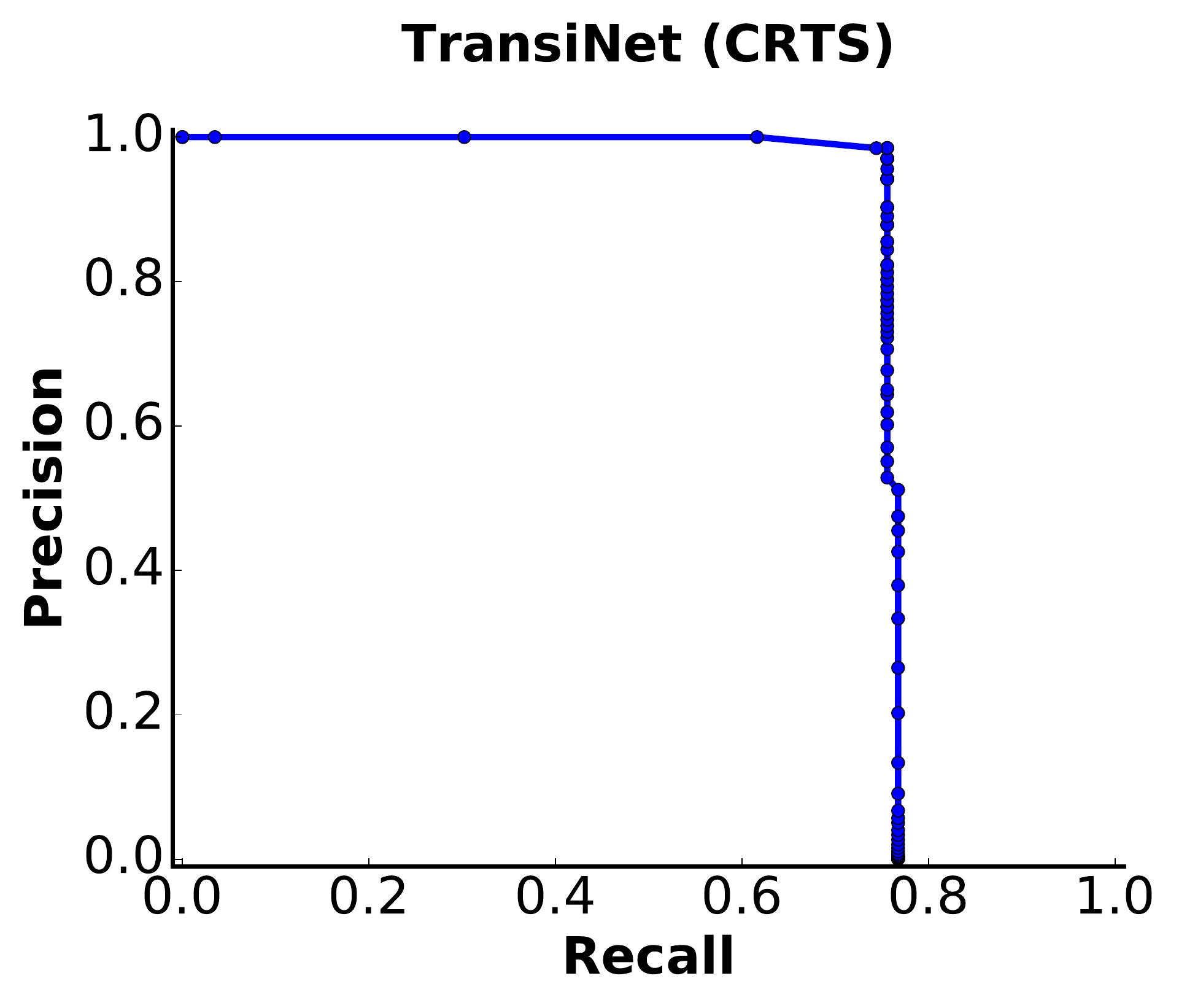}
    \includegraphics[width=0.48\linewidth]{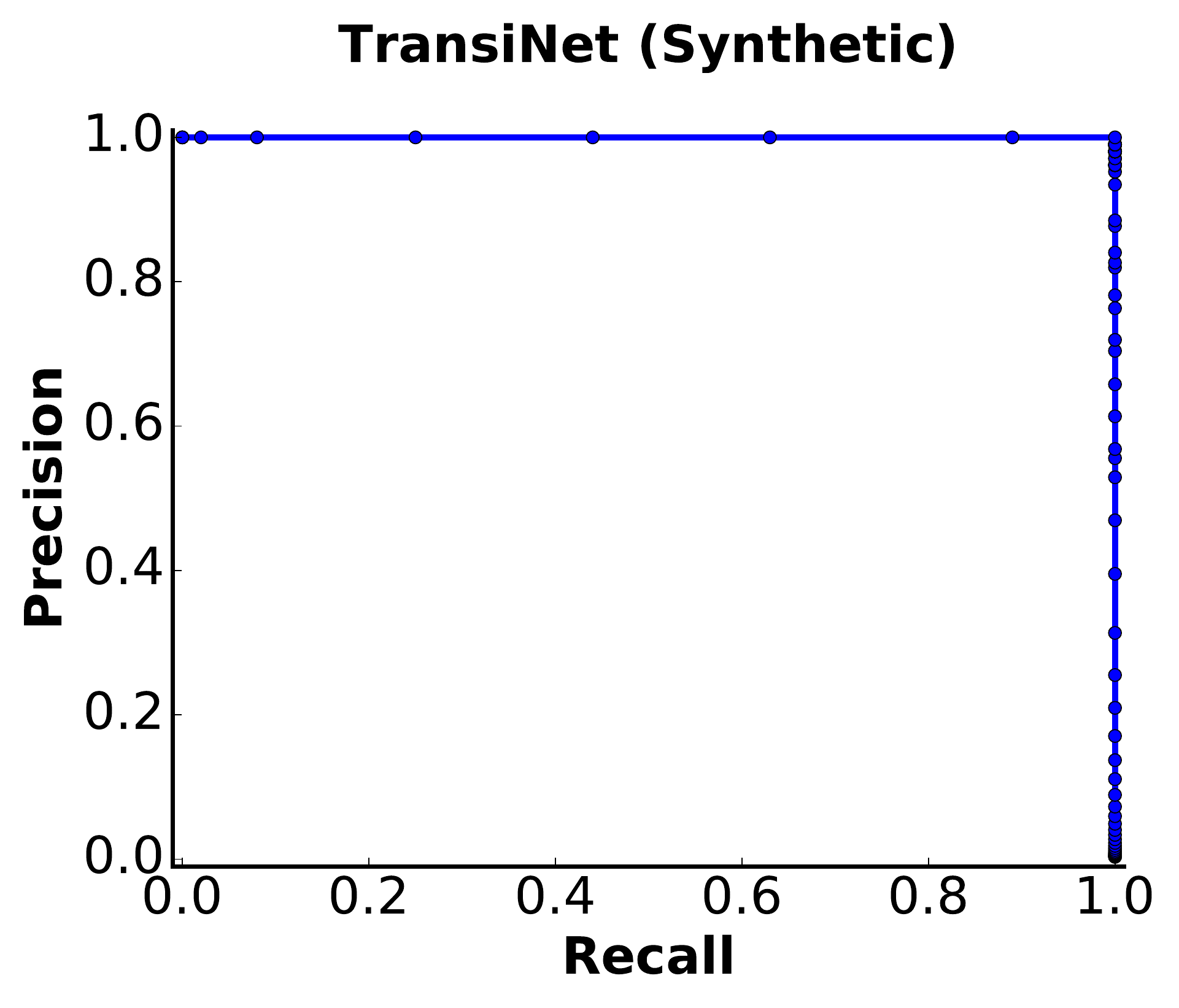}
    \includegraphics[width=0.48\linewidth]{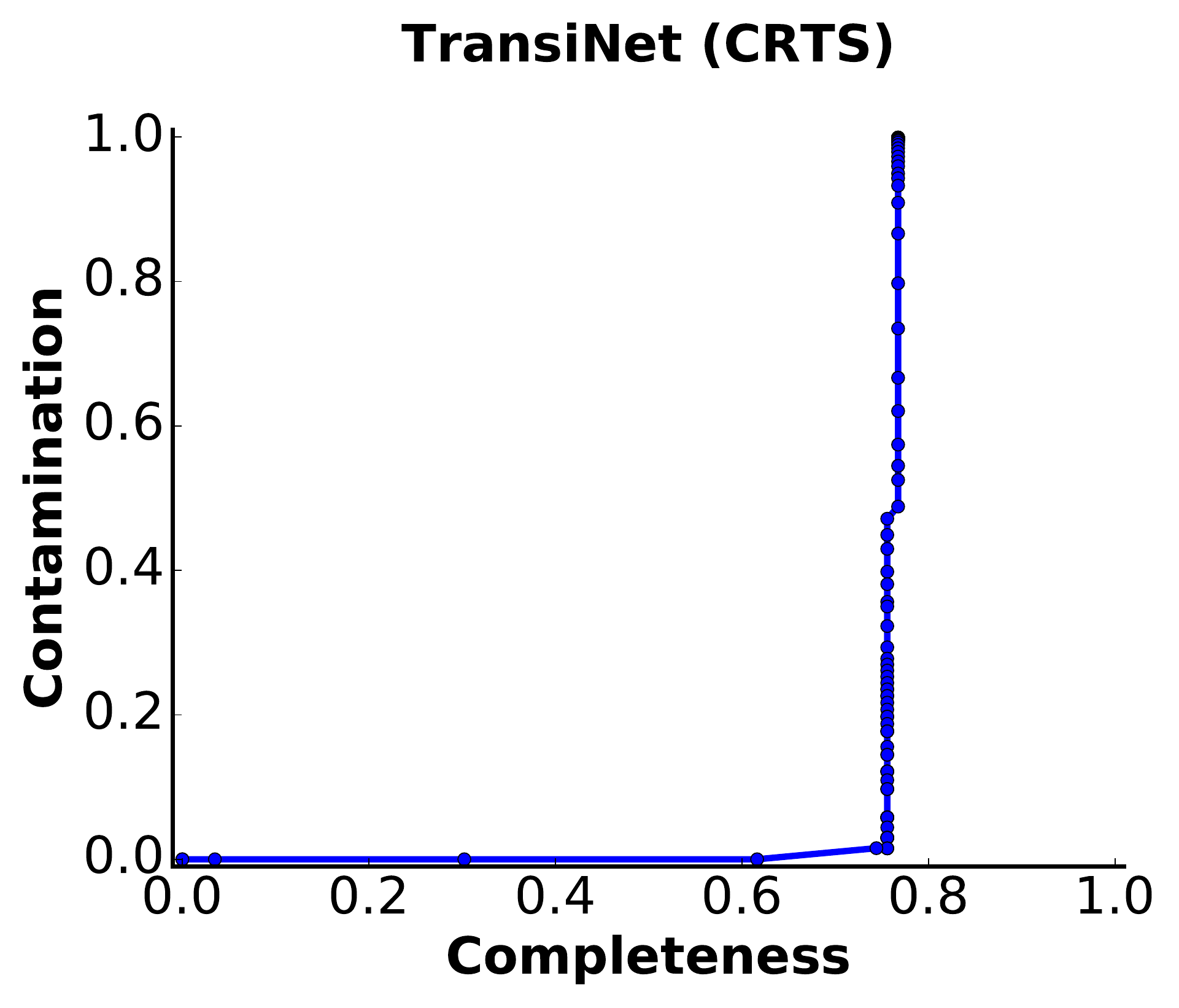}
    \includegraphics[width=0.48\linewidth]{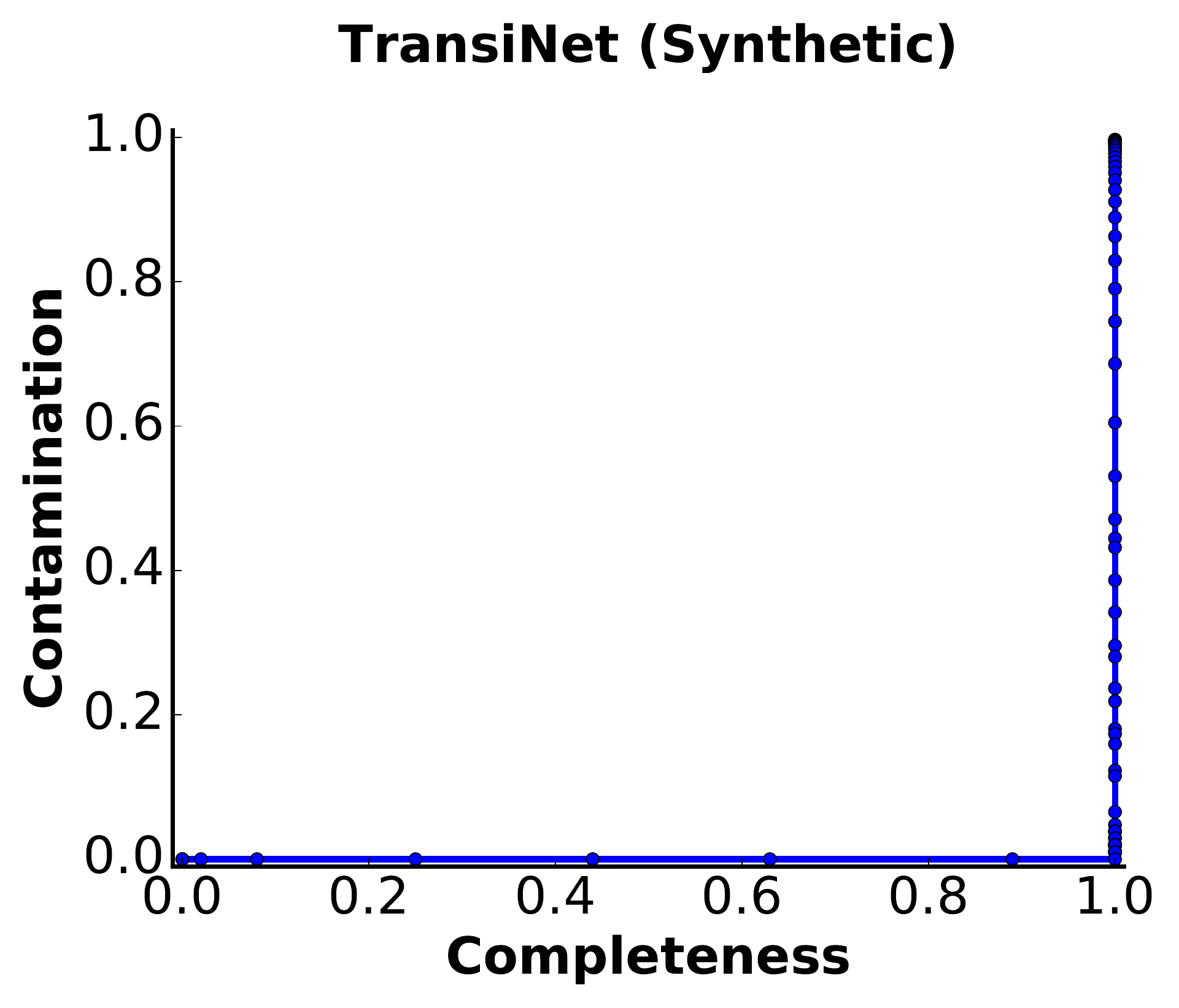}
    \includegraphics[width=0.48\linewidth]{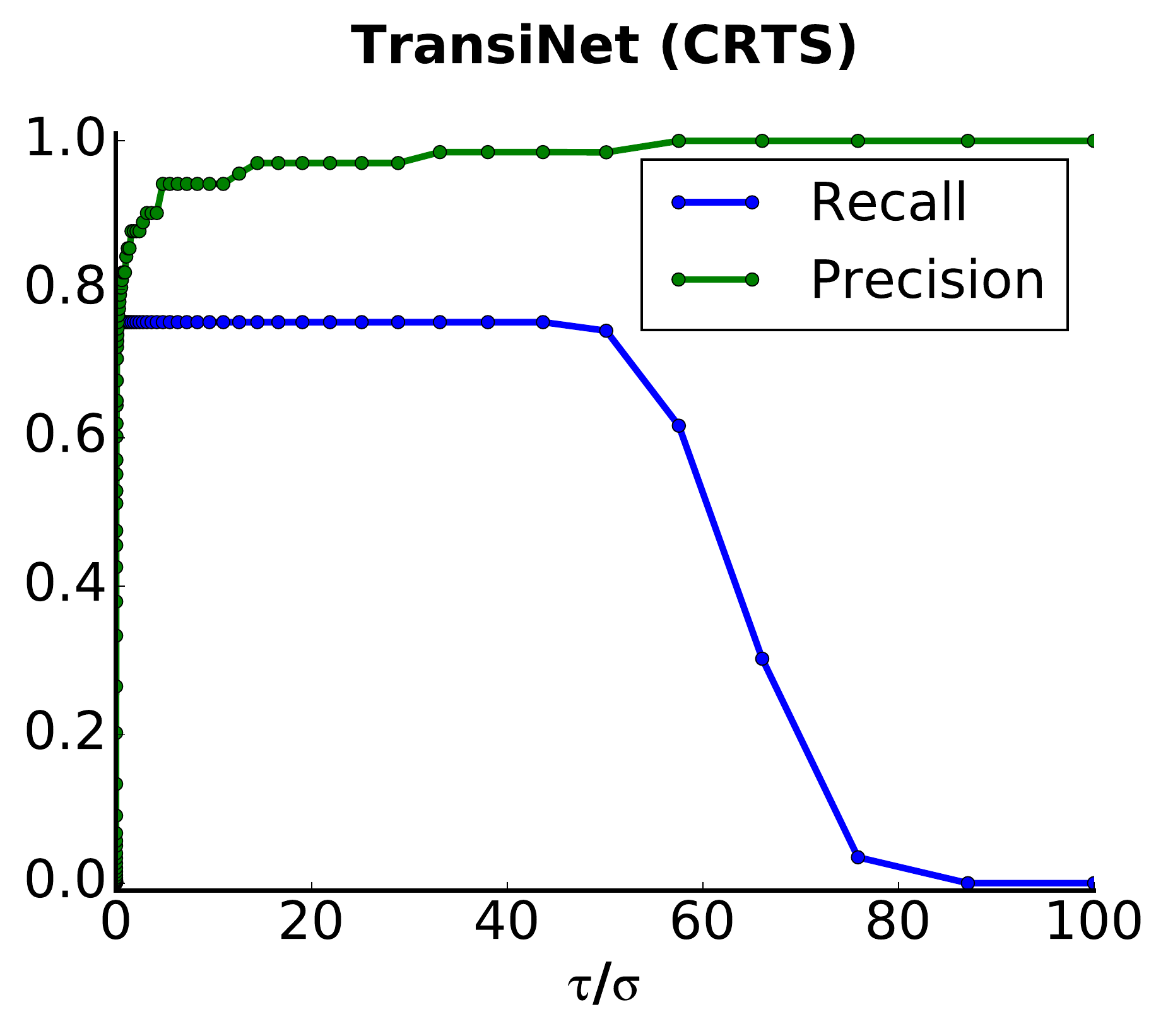}
    \includegraphics[width=0.48\linewidth]{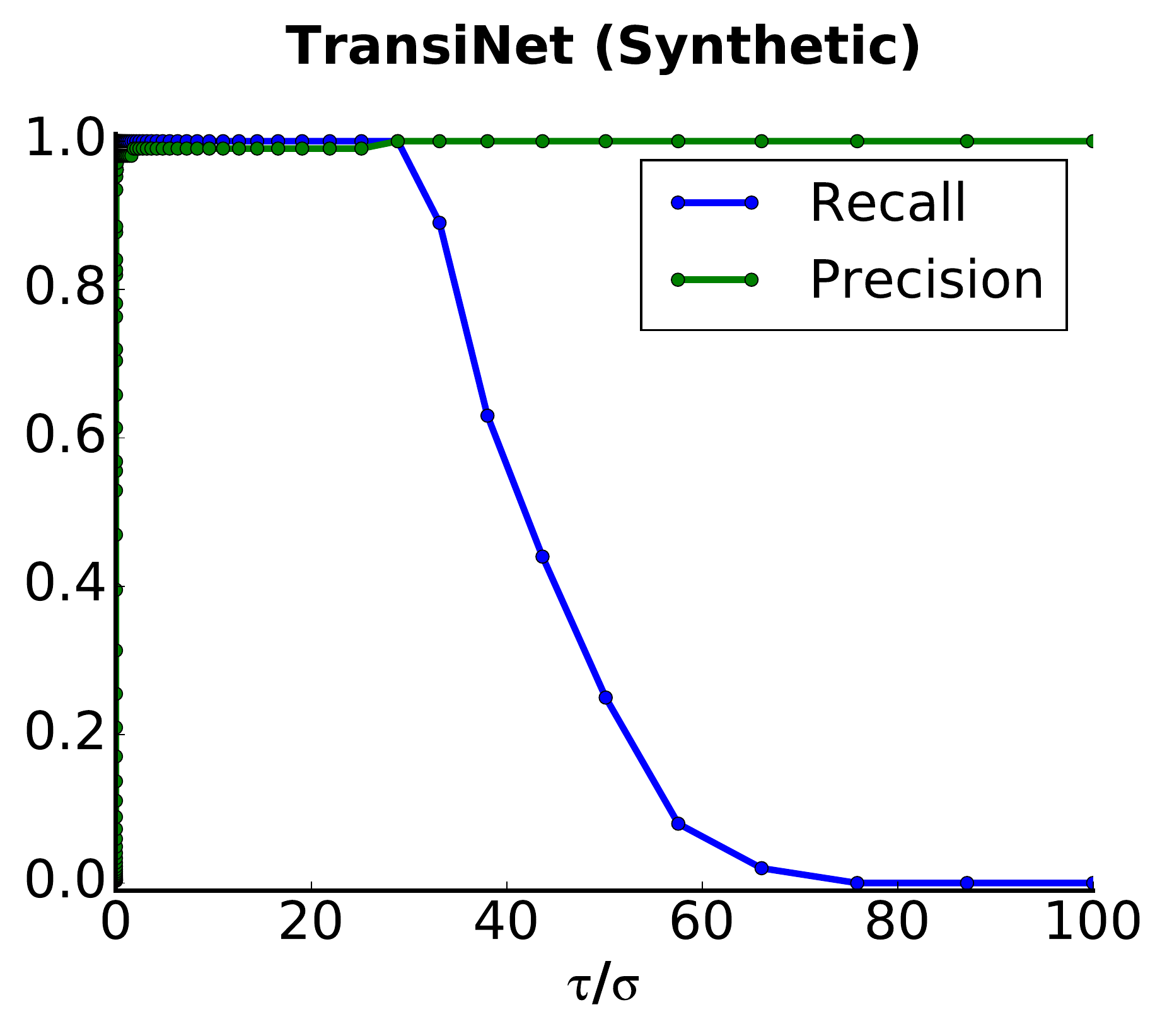}
    \includegraphics[width=0.48\linewidth]{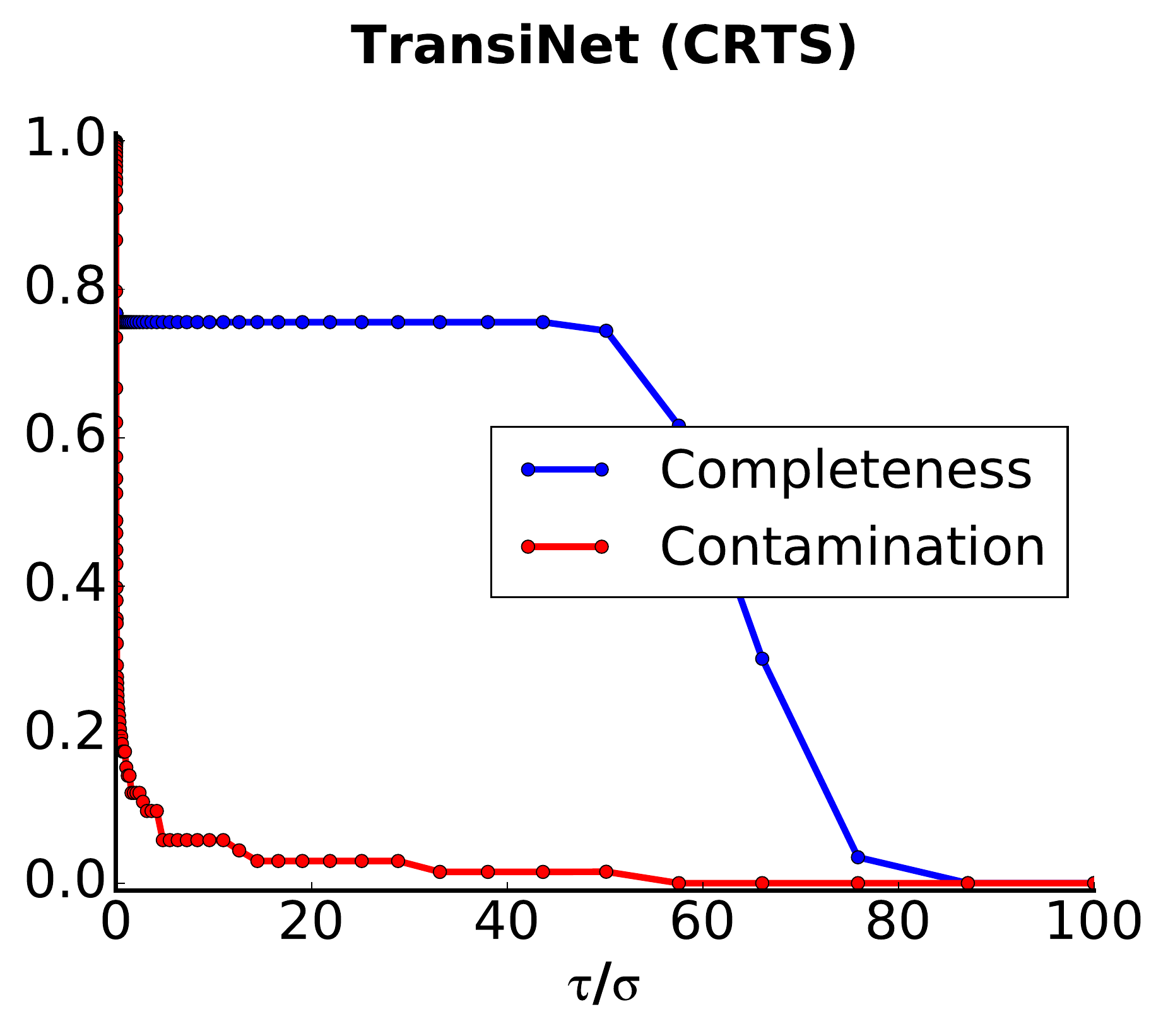}
    \includegraphics[width=0.48\linewidth]{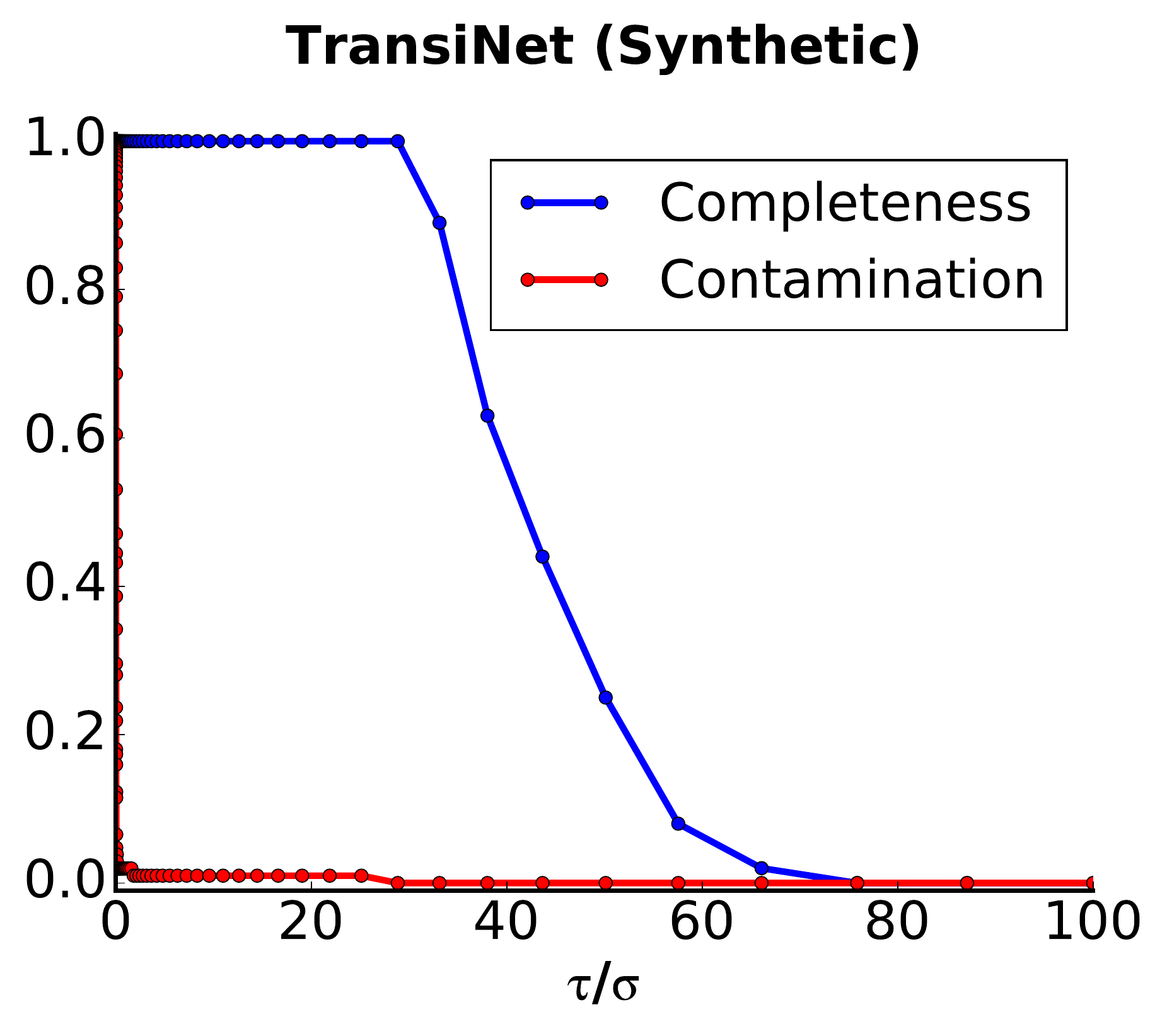}
    
  \end{center}
  \caption{
  The plots are as in Fig.~\ref{fig:precision-recall-preblanking}, but post-blanking.
The plots are now closer to ideal. 
In this scenario, for CRTS, we never go above a completeness of 80\% detections, but all those detections are clean, and the ones we miss are the really low significance ones below the blanking threshold of $0.001$. Fig.~\ref{fig:thresholding} shows a single transient field for each type at different thresholds.}
  \label{fig:precision-recall}
\end{figure}

Let $P$ be the set of all positives, i.e. the objects in $\hat{Y}$, and $G$ the set of all objects in $Y$. Then we have
\begin{equation}
	\TP = P \wedge G
\end{equation}
\noindent where $\wedge$ is used here to denote spatial intersection, such that $\TP$ is the set of objects in $P$ that have an spatial intersection with a member of G. $\TP$ is the set of \emph{True Positives}. We conversely define $\TP' = G \wedge P$ which is of the same cardinality as $\TP$ and includes the set of objects in $G$ that have been detected. Then we also have:
\begin{gather}
	\FP = P - \TP
    \\
    \FN = G - \TP'
\end{gather}
\noindent in which $\FP$ and $\FN$ stand for \emph{False Positives} and \emph{True Positives} respectively. Now we can rewrite \Cref{eq:precision_recall_eq1_1,eq:precision_recall_eq1_2} in a more compact and formal form as:
\begin{gather} 
	Precision = \frac{|\TP|} {|\TP| + |\FP|}
    \\
	Recall = \frac{|\TP|} {|\TP| + |\FN|}
\end{gather}
\noindent where $|\cdot|$ represents the cardinality of the set.
We also define \emph{completeness} and \emph{contamination} measures as follows:
\begin{gather} 
Completeness = \frac{|\TP|} {|\TP| + |\FN|} = Recall	\\
Contamination = \frac{|\FP|} {|\TP| + |\FP|} = 1 - Precision	
\end{gather}
Figs.~\ref{fig:precision-recall-preblanking} and \ref{fig:precision-recall} depict the precision-recall curves  corresponding to the two versions of TransiNet before and after blanking. Each curve is obtained by sweeping the threshold ($\tau$) in the pixel-value domain. Starting from the minimum (0), $\hat{Y}$ is set to $1$ everywhere, resulting in a 100\% recall (everything that is to be found is found) with a close-to-zero precision (too many false positives), which is equivalent to total contamination. But as we increase $\tau$, fewer pixels in $\hat{Y}$ `fire', generally resulting in a lower recall (some misses) and higher precision (far fewer contaminants) -- see Fig.~\ref{fig:thresholding}. To generate the curves we sampled 101 logarithmically-distributed values for $\tau$ from the range $[10^{-4}\sigma,100\sigma]$, where $\sigma$ is the standard deviation of the pixel values in each detection image ($y$). Also the ground truth images were binarized with a fixed threshold of $10^{-3}$. 

The sharp and irregular behavior of the curve at around 75\% of recall on the CRTS dataset, is due to the low contamination levels in the output: transients are detected with a high significance. Contaminants, if any, have a much lower intensity, and their number goes up only when one pushes for high completeness to the lower significance levels. 

\subsubsection{Relative Magnitude of the Transient}

\begin{figure}[]
  \begin{center}
    \includegraphics[width=0.90\linewidth]{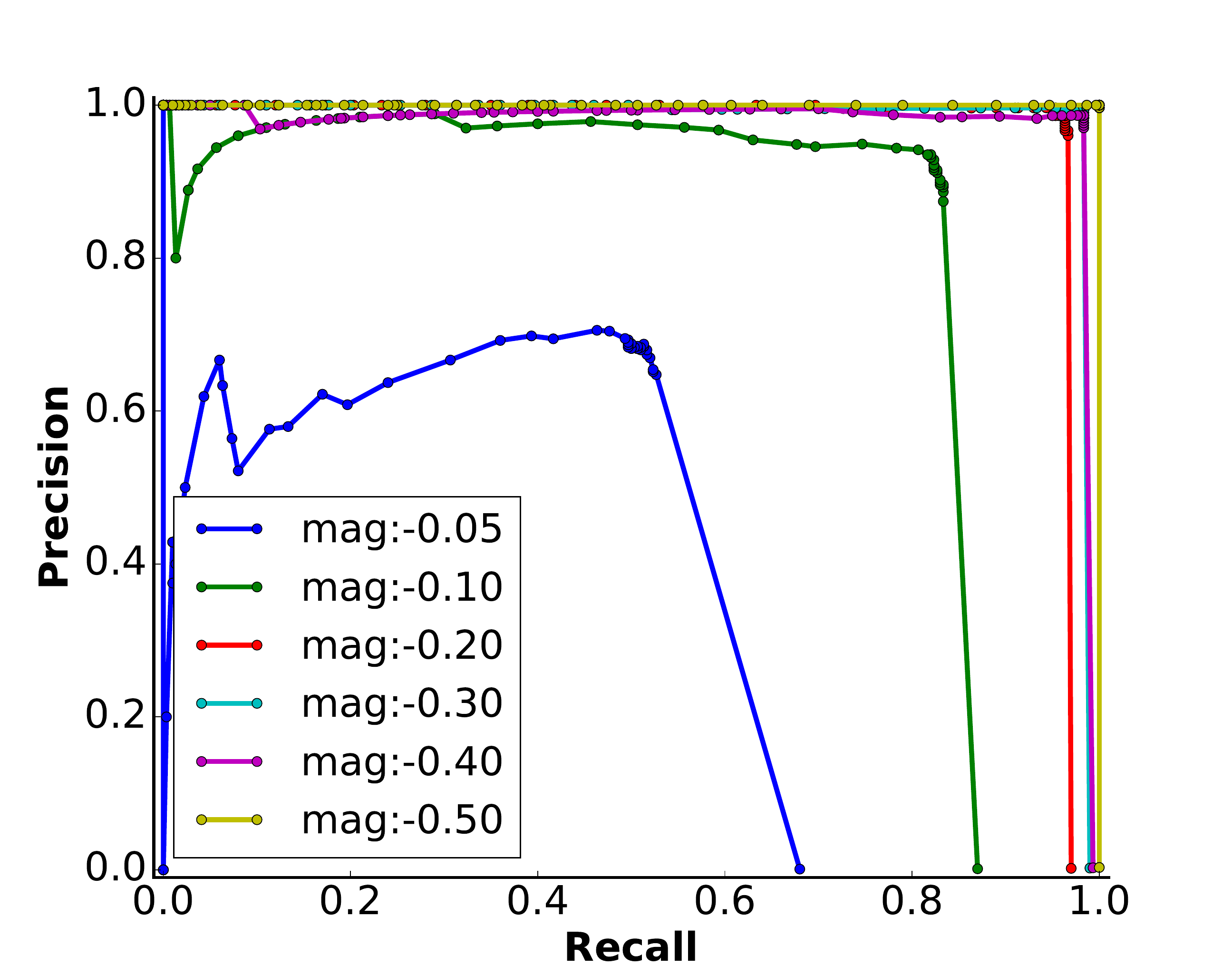}
  \end{center}
  \caption{Precision-recall curves for a range of magnitudes. These are for the synthetic transients where we had control over the relative magnitudes. The network misses more transients as the relative magnitude goes lower. This is not unexpected as the network has not seen such faint samples during training. The sharp vertical transitions reflect the clean nature of the detection images.}
  \label{fig:multimag}
\end{figure}

Thanks to the freedom in generation of synthetic samples with different parameters, we can evaluate the performance of the network with transients of different magnitudes. However, for this evaluation we use \emph{relative magnitudes}, as opposed to the absolute intensities used during training. This would make it easier to quantitatively determine the ability of the network to detect faint transients without contamination. In the future we hope to incorporate similar process during training as well.

We define the relative magnitude as the difference of magnitudes at the location of transient, with and without the transient: 
\begin{gather}
  mag_{rel} = -2.5 \log_{10} (F_{rel})\\
  F_{rel} = \frac{F_t+F_{local}}{F_{local}}
\end{gather}

\noindent where $F_{t}$ is the absolute flux of the transient, and $F_{local}$ represents the flux of the background,  before having the transient. The latter is measured inside an FWHM-sized square neighborhood around the location of the transient.

Fig.~\ref{fig:multimag} depicts the performance of the detector for several relative magnitudes, in terms of the precision-recall curve. With higher visibility, the curve approaches the ideal form. Considering that during the training phase, the network has rarely seen transients with such low magnitudes as the ones in the lower region of this experiment, it is still performing well. We expect it to gain much better results by broadening the range of simulated transient amplitudes during training.

\subsubsection{Robustness to Spatial Displacements}
\begin{figure}[]
  \begin{center}
    \includegraphics[width=0.48\linewidth]{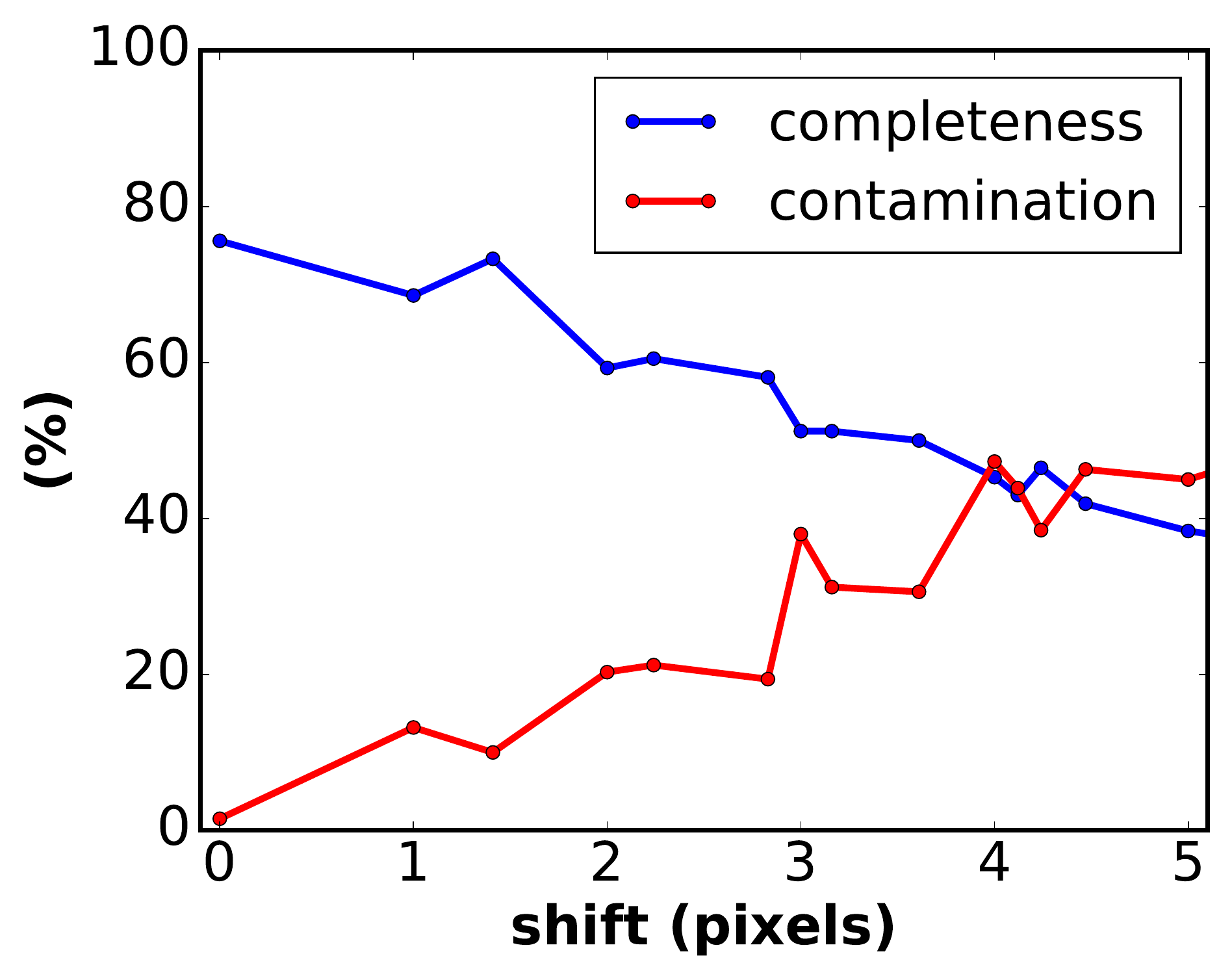}
    \includegraphics[width=0.48\linewidth]{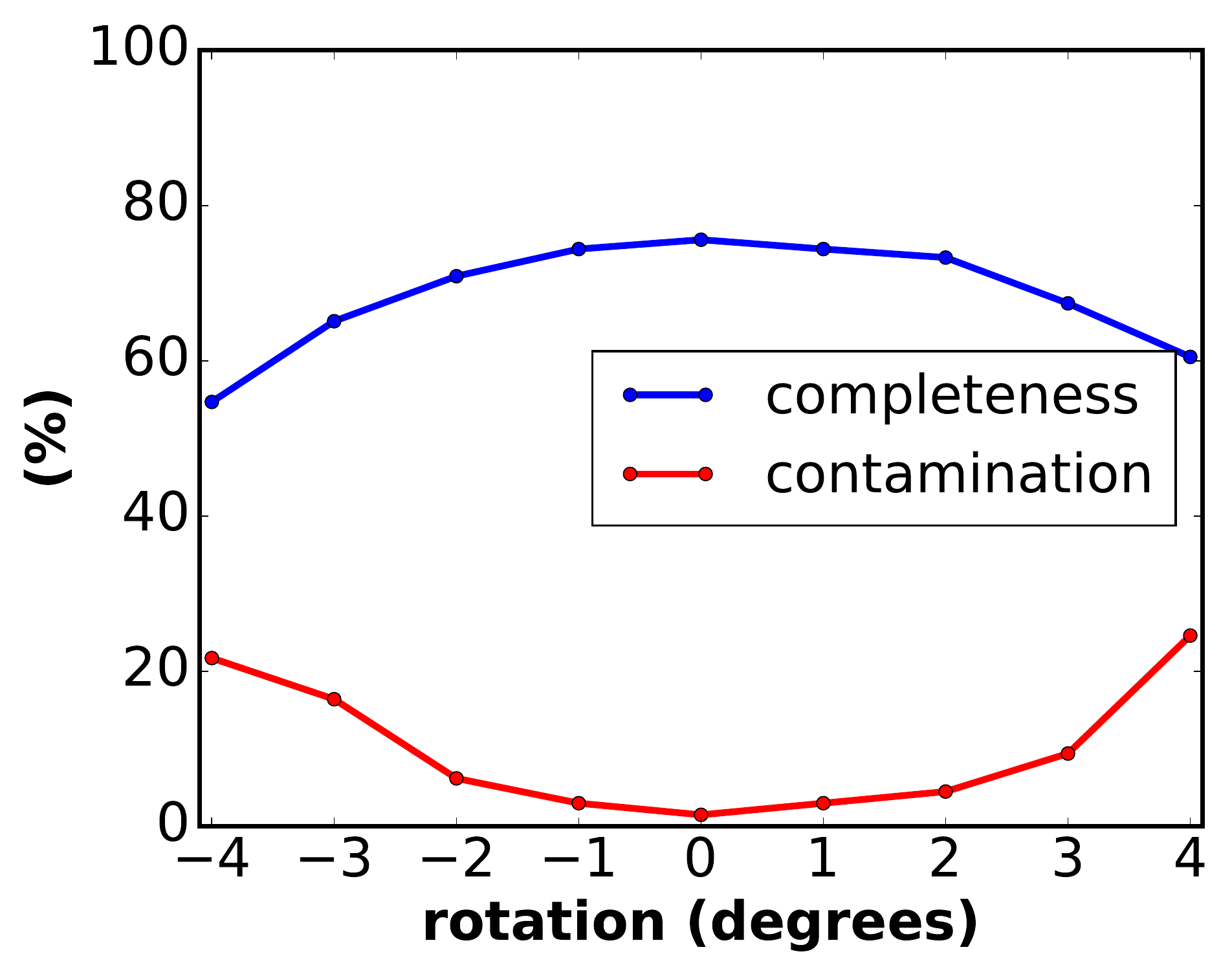}
    \includegraphics[width=0.48\linewidth]{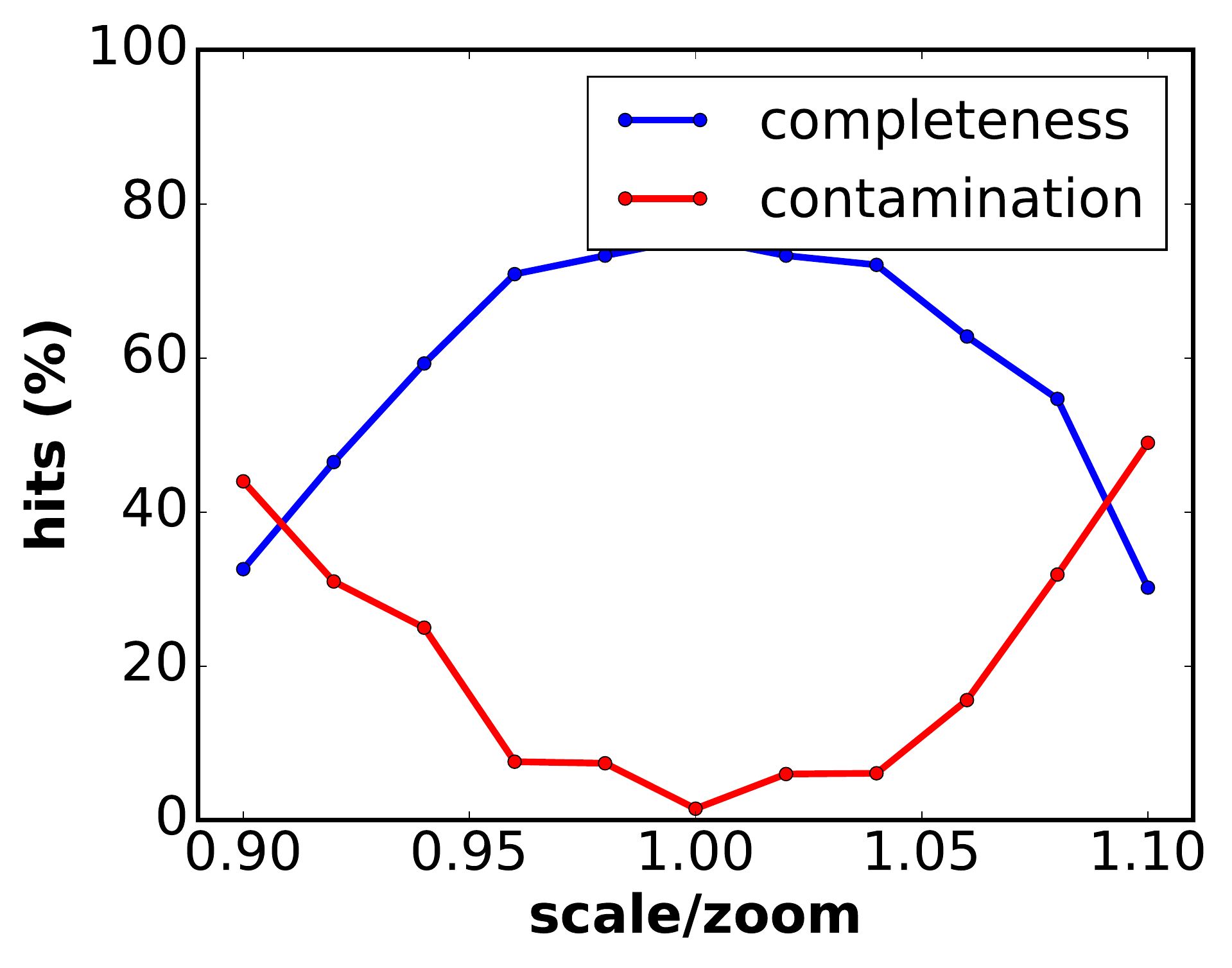}
  \end{center}
  \caption{Robustness of the network to shift (top-left), rotation (top-right), and scaling (bottom) between the reference and science images. Ideally there will be no misalignments, but some can creep in through improper WCS, or changes between runs etc. The peak detectability is close to 75\% across the board.
  }
  \label{fig:sensitivity}
\end{figure}

We analyze the robustness of the TransiNet to pairwise spatial inconsistencies between the science and reference images. That way small rotations, WCS inconsistencies etc. do not give rise to Yin-Yang like `features' and lead to artifacts. 
To this end, for a subset of image pairs, we exert manual shift, rotation and scaling to one of the images in each pair, and pass them through the network. Fig.~\ref{fig:sensitivity} depicts the results of these experiments as plots of completeness and contamination vs. the manual perturbation.

\subsubsection{Performance of the two TransiNet Networks}

Table~\ref{tab:results} summarizes the testing results for the two TransiNet networks. For new surveys one can start with the generic network, and as events become available, fine-tune the network with specific data. 

\subsection{Comparison with ZOGY}

Given the generative, and hence very different, nature of our `pipeline', it is difficult to compare it with direct image differencing pipelines. We have done our best by comparing the output of TransiNet and of ZOGY for synthetic as well as real images. We used the publicly available MATLAB version of ZOGY, and almost certainly we used ZOGY in a sub-optimal fashion. As a result this comparison should be taken only as suggestive. More direct comparisons with real data (PTF and ZTF) are planned for the near-future. Fig.~\ref{fig:results} depicts the comparison for a few of the SN Hunt transients.

\begin {table}
\begin{center}
\begin{tabular}{lrrrrrrr}
\hline
Network & Transients & TP & FP & FN & Prec. & Recall\\
\hline
Synth/Zoo & 100 & 100 & 0 & 0 & 100.0 & 100.0 \\
CRTS/SNH & 86 & 65 & 1 & 21 & 98.4 & 75.5\\
\hline 
\end{tabular}
\caption {\label{tab:results} 
Hits and misses for TransiNet for the Synthetic and SN Hunt networks. TransiNet does very well for synthetics. One reason could be that there isn't enough depth variation in the reference and science images. But for CRTS too the output is very clean for the recall of 76\% that it achieves. The lower (than perfect) recall could be due to a smaller sample, larger pixels, large shifts in some of the cutouts etc. Fine tuning with more data can improve performance further. The fixed thresholds used for the synthetic and SN Hunt networks are 40 and 20 respectively.
}  
\end{center}
\end {table}

Both pipelines could be run in parallel to choose an ideal set of transients, since the overhead of TransiNet is minuscule.

\section{Future Directions} 
We have shown how transients can be effectively detected using TransiNet. In using the two networks we described, one with the Kaggle zoo images, and another with CRTS, we cover all broad aspects required, and yet for this method to work with any specific project, e.g. ZTF, appropriate tweaks will be needed, in particular labeled examples from image differencing generated by that survey. Also, the assumptions during simulations can be improved upon by such examples. Using labeled sets from surveys accessible to us is definitely the next step. Since the method works on the large pixels that CRTS has, we are confident that such experiments will improve the performance of TransiNet.

The current version produces convolved transients to match the shape and PSF of the science image. One can modify the network to produce just the transient location and leave the determination of other properties to the original science and reference images as they contain more quantitative detail.

Further the network could be tweaked to find variable sources too. But for that a much better labeled non-binary training set will be needed. In the same manner, it can also be trained to look for drop-outs, objects that have vanished in recent science images but were present in the corresponding reference images. This is in fact an inverse of the transients problem, and somewhat easier to do.

In terms of reducing the number of contaminants even further, one can provide as input not just the pair of science and reference images, but also pairs of the rotated (by 90, 180, 270 degrees) and flipped (about x- and y-axes) versions. The expectation is that the transient will still be detected (perhaps with a slightly different peak, extent), but the weak contaminants, at least those that were possibly conjured by the weights inside the network, will be gone (perhaps replaced by other -- similarly weak ones -- at a different location), and the averaging of the detections from the set will leave just the real transient.

Another way to eliminate inhomogeneities in network weights is to test it with image pairs without any transients. While most image pairs do not have any transient except in a small number of pixels, such a test can help streamline the network better.

In order to detect multiple transients, one could cut the image in to smaller parts and provide these submimages for detection. Another possibility is to mask the `best' transient, and rerun the pipeline to look for more transients iteratively until none is left. An easier fix is to train the network for larger images, and for multiple transients in each image pair.

Another way to improve the speed of the network is to experiment with the architecture, and if possible obtain a more lightweight network with a smaller footprint that performs equally well. Finally, the current network used jpegs with limited dynamic range as inputs. Using non-lossy FITS images should improve performance of the network.
\section{Conclusions}
\label{Sec:discussions}
We have introduced a generative method based on convolutional neural networks for image subtraction to detect transients. It is superior to other methods as it has a higher completeness at lower thresholds, and at the same time has fewer contaminants. Once the training is done with appropriate labeled datasets, execution on individual images is fast. It can operate on images of any size (after appropriate training), and can be easily incorporated in to real-time pipelines. While we have not explicitly tested the method on high-density fields (e.g. closer to the plane of the Galaxy) it will be possible to get good performance once a corresponding labeled dataset is used for training. We hope surveys like ZTF, LSST as well as those with larger pixels like ASAS-SN \citep{Shappee2014}, Evryscope \citep{Law2015} etc. adapt and adopt the method. It is also possible to extend the method to other wavelengths like radio and use for surveys including SKA and its path-finders.


\section*{Acknowledgements}

AM was supported in part by the NSF grants AST-0909182, AST-1313422, AST-1413600, and AST-1518308, and by the Ajax Foundation.




\bibliographystyle{mnras}
\bibliography{transient_hunting} 





\bsp	
\label{lastpage}
\end{document}